\documentclass[reprint,onecolumn,longbibliography,eqsecnum,floats,aps,amsmath,amssymb,nofootinbib,prd,showpacs,notitlepage]{revtex4-1}
\usepackage[a4paper,top=3cm,bottom=3cm,left=2cm,right=2cm,marginparwidth=1.75cm]{geometry}
\usepackage[english]{babel}
\usepackage[utf8]{inputenc}
\usepackage{amsmath}
\usepackage{amssymb}
\usepackage{lmodern}
\usepackage{nccmath}
\usepackage{graphicx}
\usepackage{wrapfig}
\usepackage{enumerate}
\usepackage{geometry}
\usepackage{hyperref}
\usepackage{braket}
\usepackage{float}
\usepackage{subfig}
\usepackage{dsfont}
\DeclareMathOperator*{\Tr}{Tr}

\newcommand{\rd}{{\rm d}}
\newcommand{\re}{\mathbb{R}}
\newcommand{\Hil}{\mathcal{H}}
\DeclareMathOperator*{\sgn}{sgn}
\newcommand{\ub}[1]{\underline{#1}}
\newcommand{\Weyl}[1]{{\boldsymbol{:}\! #1 \!\boldsymbol{:}}}

\begin{document}

\title{Regularizations and quantum dynamics in loop quantum cosmology}

\author{Maciej Kowalczyk}
	\email{maciej.kowalczyk@uwr.edu.pl}
	\affiliation{Institute for Theoretical Physics, Faculty of Physics and Astronomy, University of Wroc{\l}aw, pl. M. Borna 9, 50-204  Wroc{\l}aw, Poland}
\author{Tomasz Paw{\l}owski}
	\email{tomasz.pawlowski@uwr.edu.pl}
	\affiliation{Institute for Theoretical Physics, Faculty of Physics and Astronomy, University of Wroc{\l}aw, pl. M. Borna 9, 50-204  Wroc{\l}aw, Poland}

\begin{abstract}
  One of critical components of Loop Quantum Gravity (LQG) and Cosmology (LQC) -- Thiemann regularization procedure is non-unique. Different choices of particular prescriptions lead to models which differ in both mathematical structure and physical predictions. Here we briefly recall a set of such prescriptions proposed in the literature in context of isotropic LQC on the example of a flat universe with massless scalar matter content. For the one least investigated so far, further called Yang-Ding-Ma prescription, a detailed analysis of its mathematical structure and resulting quantum dynamics is performed, confirming and extending the results obtained so far by phenomenological methods. In order to probe the dynamics, a relatively robust method (working in the approximation of the macroscopic universe) of evaluating quantum trajectories is devised. Said method is a variant of a semiclassical treatment that allows to express the trajectories analytically as function of internal clock and a set of certain central moments -- constants of motion encoding quantum corrections up to arbitrary order in systematic manner. As a test of method's robustness analogous evaluation of the quantum trajectory in volume is performed for those of other prescriptions, for which it is applicable. The limitations of the treatment are further briefly discussed.
\end{abstract}


\maketitle

\section{Introduction}

Loop Quantum Gravity (LQG) \cite{Thiemann:2007pyv,Rovelli:2004tv} --one of the more popular approaches to build a consistent quantum description of both spacetime geometry and matter-- has undergone significant progress in recent decades. Thanks to implementing an unorthodox quantization procedure known as polymer quantization \cite{Thiemann:1996aw}, which sidesteps the quantum mechanics uniqueness theorems, it offers hope for predicting new phenomena in physical processes involving high energies and gravitational fields. These in turn might provide an explanation for newly observed features in precise cosmological measurements \cite{Agullo:2012sh,*Agullo:2014ica,deBlas:2016puz,*CastelloGomar:2017kbo} or offer a solution for black hole information loss paradox \cite{Ashtekar:2005cj,Haggard:2014rza,*Bianchi:2018mml}. On one hand, the applied quantization significantly alters the classical properties of geometry, for example predicting the discrete nature of geometry quantities (i.e. discrete spectra of operators measuring areas and volumes) offering a solid prospect for a new physics. On the other hand, it poses a series of hard challenges. 

One of them is an enormous complication of its mathematical structure, which makes extracting physical predictions enormously challenging on the technical level. For example, while there exist a set of solid frameworks allowing to define the dynamics of quantum space \cite{Giesel:2007wn,Domagala:2010bm,Husain:2011tk}, actual evaluations of the dynamical evolution are so far restricted to ultra-simple unphysical states \cite{Zhang:2019dgi} and phenomenological semiclassical approaches \cite{Dapor:2017rwv,*Dapor:2020jvc}. To overcome this problem, researchers turn to various simplifications of LQG. One of the most known and most radical is Loop Quantum Cosmology (LQC)\cite{Bojowald:2002gz,Ashtekar:2008zu,Ashtekar:2011ni} constructed via application of the mathematical methodology of LQG to descriptions of highly symmetric spacetimes, usually cosmological ones. This framework is sufficiently simple to allow for making precise dynamical predictions on the genuine quantum level in the case of simplest models (homogeneous spacetimes) \cite{Ashtekar:2006rx,Ashtekar:2006wn} and further extrapolate the genuine quantum results via phenomenological methods to more realistic scenarios (for example cosmological perturbations). This in turn allowed to find some predictions regarding for example spectra of cosmic microwave background \cite{Agullo:2012sh,*Agullo:2014ica,deBlas:2016puz} or yet unobserved gravitational wave background \cite{Ramirez:2018dxe}. However, these results, obtained with various levels of rigour, are not predictions of LQG per se. Instead, they are obtained via a framework that is to a high degree independent from it and reconciling the cosmological sector of LQG with LQC is an ongoing task.

The second challenge is the very same mechanism, that offered hope for yet undiscovered new physical phenomena -- the sidestepping of the quantum mechanics uniqueness theorems, which poses a danger that various choices made to resolve the quantization ambiguities may lead to different dynamical predictions. One of the principal sources of such ambiguities is the so-called \emph{Thiemann regularization} -- a process of rewriting the local differential quantities (like curvature) via extended ones (holonomies and fluxes), which are then being quantized \cite{Thiemann:2007zz}. While probing the consequences of such ambiguities in full LQG is extremely difficult due to insufficient control over the dynamics, one can probe them in LQC, which (as implementing the very same methodology) is a direct analog of LQG in simpler settings. There indeed, the analysis of the dynamics of simplest (isotropic) models have shown, that different choices (regularization prescriptions), while recovering the classical relativity in low energy limit, do provide different predictions in a high energy "near classical singularity" regime, often leading to significant qualitative changes in some aspects (see for example a comparizon of the results of \cite{Ashtekar:2006wn} with \cite{Assanioussi:2018hee}). 

In the context of the simplest isotropic LQC, so far three regularization prescriptions have been proposed and studied in the literature. They all differ in a way one reexpressed the so-called Lorentzian term of the Hamiltonian constraint \cite{Ashtekar:2006uz} in terms of holonomies and volume operator. The earliest proposal (further referred to as the \emph{mainstream LQC}) \cite{Ashtekar:2006wn} is based on splitting said term onto a linear combination of the spatial Ricci scalar and the part of the Hamiltonian constraint proportional to the curvature of Ashtekar connection (the so-called \emph{Euclidean part}). The second one \cite{Yang:2009fp}, implements strictly the algorithm of reexpressing the exterior curvature originally proposed by Thiemann for full LQG \cite{Thiemann:1997rt}. The mathematical structure and the physics of quantum models resulting from these choices have been investigated in detail. The third one \cite{Yang:2009fp} is based on a simple 1st order approximation of the curvature operator in terms of the Ashtekar connection. Its studies so far involved only phenomenological studies of the dynamics (via $0$th order effective dynamics \cite{Li:2019ipm}) and an analysis of the stability of the eigenvalue problem \cite{Saini:2019tem}. The conclusions of the latter indicated, that, unlike for the previous prescriptions, the physical Hilbert space, if exists, is spanned by some unidentified proper subspace of the solutions to that problem. Thus, in order to establish the validity of this prescription a detailed analysis of the resulting quantum model is necessary. We perform this analysis here, probing in particular the selfadjointness properties of the evolution generator and explicitly constructing the physical Hilbert space and physically meaningful observables. Furthermore, the quantum trajectories of the system are determined analytically by a newly introduced quite general recipe, that for physically relevant states permits to include quantum corrections of arbitrary order.

It is also worth mentioning, that generalizations on more fundamental level have also been considered in the literature. In particular, consequences of modifying the original reduced holonomy-flux algebra, using the so called \emph{flux-covariant holonomies} \cite{Liegener:2019ymd} (see also further application in black hole models \cite{Giesel:2021dug}) have been probed. While the preliminary studies suggested significant modifications to the dynamics, the subsequent rigorous studies in \cite{Kaminski:2020wbg} have shown, that the new algebra has \emph{the same} Poisson structure as the original one and the dynamical predictions of the model are the same as those of mainstream LQC.

The paper is organized as follows: 
We begin by briefly recalling the LQC quantization procedure in the context of the simplest cosmological model --flat FLRW universe-- in sec.~\ref{sec:model}, in particular outlining the constructions and main properties of three regularization prescriptions discussed in the literature. 
Next, in sec.~\ref{sec:mLQC2} we will study in detail the properties of the quantum evolution generator (an equivalent of a quantum Hamiltonian operator) constructed with the use of the third regularization prescription listed in the previous paragraph. In particular, the selfadjointness of the evolution generator (thus the uniqueness of the quantum evolution will be probed through the deficiency analysis in sec.~\ref{sec:theta-defic}. Subsequently, the spectral properties of the evolution generator will be studied and an analog of the energy eigenbasis of the geometry Hilbert space will be explicitly contructed. These results will then be used in sec.~\ref{sec:Hphys} to construct physical Hilbert space and physical observables. 
Subsequently, sec.~\ref{sec:trajectories} will be dedicated to determining the quantum trajectories -- in this case volume (and its variance) as a function of the matter clock. For that, a relatively general method of evaluating the expectation value of the relevant observables analytically will be presented and used. Said method, under quite natural physical assumption -- a large ``energy'' (momentum of the clock field) of the universe, will allow to capture quantum corrections up to arbitrary order. For completeness, we will also apply our method to derive the quantum trajectories for the remaining two prescriptions and the geometrodynamical analog of the studied model, thus giving higher order quantum corrections to the results obtained via the ($0$th order) effective dynamics.
Finally, we will conclude in sec.~\ref{sec:conclusions} with a discussion of the results and the perspectives on applying presented trajectory probing method in a wider set of models.

\section{Regularizations in LQC - flat FLRW Universe}
\label{sec:model}

Let us start with recalling the basic structure of the example model to be used for our studies -- the LQC quantized flat isotropic FRLW universe admitting massless scalar field as a matter content. This model (and its treatment) has been already extensively presented in the literature (see for example  \cite{Ashtekar:2006wn,Ashtekar:2007em,Assanioussi:2018hee}). Being a quantization of a classical theory with constraints, it follows the so-called Dirac quantization programme, where the quantum theory is built in steps: the kinematical one (ignoring the constraints) and the implementation of the constraints in order to construct the physical one. We start with kinematics.

\subsection{Classical description and kinematics}
\label{sec:class_kin}

Here we follow directly \cite{Ashtekar:2003hd} with further improvements proposed in \cite{Ashtekar:2006wn} and \cite{Ashtekar:2009vc}.
The point of departure is the restriction of the canonical formulation of general relativity in terms of Ashtekar variables \cite{BarberoG:1994eia} to isotropic spacetimes. First, we choose foliation by homogeneity surfaces and introduce an auxiliary structure - a constant (in comoving coordinates) orthonormal spatial triad ${}^oe^a_i$ and its dual cotriad ${}^o\omega^i_a$, which in turn forms a fiducial metric 
${}^oq_{ab} = \delta_{ij}{}^o\omega^i_a {}^o\omega^j_b$. By partial gauge fixing we now can express the physical metric in terms of that fiducial structure, lapse function $N(t)$ and the scale factor $a(t)$ via $ds^2 = -N^2(t)\rd t^2 + a^2(t){}^oq$. 
Next, we introduce Ashtekar variables: densitized triad $E^a_i$ and connection $A^i_a$, further fixing the (still partial) gauge by requiring that they are proportional to 
${}^oe^a_i\tau^i$ and ${}^o\omega^a_i\tau^i$ respectively\footnote{The matrices $\tau^i$ are proportional to Pauli matrices $\tau^i=(i/2)\sigma^i$.}. To express the relation precisely we introduce a pair of global variables $(v,b)$ related with a scale factor $a(t)$ and the Hubble parameter $H_r(t)$ as follows
\begin{subequations}\label{eq:vb-def}\begin{align}
  |v| &= \left(2\pi\gamma G\hbar\sqrt{\Delta}\right)^{-1} a^3(t) 
  =: \alpha^{-1} a^3(t) , &
  b &= \gamma\sqrt{\Delta}H_r ,
  \tag{\ref{eq:vb-def}}
\end{align}\end{subequations}
where $\gamma$ is the Barbero-Immirzi parameter\footnote{In the actual calculations further in the article we will use the value determined from black hole entropy counting \cite{Ashtekar:2004eh} equaling approximately $\gamma\approx 0.2375...$ as derived in \cite{Domagala:2004jt,Meissner:2004ju}.} and $\Delta$ is the so-called \emph{area gap} of LQC\footnote{The area gap is chosen to equal twice the lowest nonvanishing eigenvalue of the area operator in LQG. For the reasoning behind this phenomenological input see for example \cite{Ashtekar:2009vc,Pawlowski:2014nfa}.}. The Poisson bracket between these variables equals then
\begin{equation}\label{eq:vb-Poisson}
  \{ b , v \} = \frac{2}{\hbar} ,
\end{equation}
and the Ashtekar variables can be written down in the following precise form
\begin{subequations}\label{eq:EA-iso}\begin{align}
  E^a_i &= \left(\frac{\alpha v}{V_o}\right)^{\frac{2}{3}} {}^oe^a_i  , 
  &
  A^i_a &= 6\pi\gamma G\hbar\, b \left(\frac{v}{\alpha^2V_o}\right)^{\frac{1}{3}} {}^o\omega^i_a  , \tag{\ref{eq:EA-iso}}
\end{align}\end{subequations}
where $V_o$ is the volume (with respect to ${}^oq$) of the so-called \emph{fiducial cell} (denoted further as $\mathcal{V}$) -- a certain compact region of the universe used to regulate otherwise infinite integrals over homogeneity surfaces.
At this moment the choice of variables appears to be rather cumbersome, however, it is in fact tailored to the procedure of Thiemann regularization further in, where the presented choice simplifies the structure of operators significantly.

The symmetries and partial gauge fixing drastically simplify the algebra of constraints normally featured in the triad formulation of GR -- the only constraint, that is not automatically satisfied, is the Hamiltonian (scalar) one. In chosen variables, it takes the form
 \begin{equation} 
H_{g}=H^E-2(1+\gamma^2)T ,
\label{eq:Hg}
\end{equation}
where $H^E$ and $T$ denote the so-called \emph{Euclidean} and \emph{Lorentzian} part respectively \cite{Ashtekar:2011ni,Thiemann:2007zz}:
\begin{align}
H^E &= 
  \frac{1}{2\kappa}\int_\mathcal{V}\rd^3x\epsilon_{ijk}\frac{E^{ai} E^{bj}}{\sqrt{\det(h)}}F^k_{ab} \ ,
 &
T &= \frac{1}{2\kappa}\int_\mathcal{V}\rd^3x\frac{E^{ai} E^{bj}}{\sqrt{\det(h)}}K^j_{[a}K^i_{b]} \ ,
\label{a1w}
\end{align}
where $F^k_{ab}$ is the curvature of Ashtekar connection $A^i_a$ and $K^i_a$ encodes the exterior curvature
\begin{subequations}\label{eq:def-FK}\begin{align}
  F^k_{ab} &= \partial_a A^k_b-\partial_b A^k_a+\epsilon_{kij}A^i_aA^j_b \ ,
  &
  K^i_a &= K_a{}^b {}^o\omega^i_b \ . \tag{\ref{eq:def-FK}}
\end{align}\end{subequations}

The quantization of the geometry degrees of freedom on the kinematical follows that of \cite{Ashtekar:2003hd}. The kinematical Hilbert space is the space of square summable functions on the Bohr compactification of the real line with the Haar measure $\rd\mu$ and is spanned by a basis formed out of almost periodic functions. The algebra of basic objects promoted to operators --a restriction of the holonomy-flux algebra-- consists of $(i)$ holonomies of $A^i_a$ along straight lines generated by ${}^oe^a_i$ and $(ii)$ fluxes of $E^i_a$ across unit squares (which for the isotropic geometries is sufficient to separate the points on the phase space). The eigenstates of the fluxes $(ii)$ form a convenient basis $\{|v\rangle\}_{v\in\mathbb{R}}$, while the holonomies $(i)$ can be expressed in terms of shift operators
\begin{subequations}\label{intv}\begin{align}
  \hat{p}\ket {v} &:= \sgn(v)\hat{V}^{\frac{2}{3}}\ket {v} \ , 
  &
  \hat{V}\ket v &=\alpha |v| \ket v \ ,
  &
  \hat{N}\ket {v}  &:= \widehat{e^{\frac{ib}{2}}}\ket{v} = \ket {v+1} \ ,
  \tag{\ref{intv}}
\end{align}\end{subequations}
where the operator $\hat{V}$ corresponds to the (physical) volume of the fiducial cell. The scalar product on the Hilbert space is the discrete one 
\begin{equation}\label{eq:ip}
  \braket{v|v'}=\delta_{v,v'} \ ,
\end{equation}

The matter degrees of freedom, in our case corresponding to a homogeneous massless scalar field (minimally coupled to gravity), are represented by a canonical pair:
field $\phi$ and its canonical momentum $p_{\phi}$ such that $\{\phi,p_{\phi}\}=1$.
Its coupling contributes to the Hamiltonian constraint as an additive term
\begin{subequations}\label{eq:C_tot}\begin{align}
  C_{{\rm tot}} &= H + C^{\prime}_{\phi} \ ,
  &
  C^{\prime}_{\phi}= \frac{p^2_{\phi}}{2V} \ ,
  \tag{\ref{eq:C_tot}}
\end{align}\end{subequations}
where $H$ is given by \eqref{eq:Hg}.

Specified matter field is quantized with standard textbook methods of quantum mechanics (using Schr\"odinger representation). The Hilbert space is the standard Lebesque one and the canonical variables are promoted to standard multiplication and derivative operator respectively. Finally, the total kinematical Hilbert space is the product
\begin{equation}
  \mathcal{H}_{{\rm kin}} = \mathcal{H}_{{\rm gr}} \otimes \mathcal{H}_{\phi}
  := \Sigma^2(\bar{\mathbb{R}},\rd\mu) \otimes L^2(\mathbb{R},\rd\phi) \ . 
\end{equation}

Having at our disposal kinematical Hilbert space and the set of fundamental operators we can now proceed with the next step of the Dirac program -- expressing the constraint as an operator.

\subsection{Constraints in LQC models}
\label{sec:constraints}

As a first step in constructing the quantum counterpart of \eqref{eq:C_tot} we note that as a constraint it can be multiplied by a lapse function without changing the resulting physics. A particularly convenient choice is (following \cite{Ashtekar:2007em}) selecting $N=2V$ as then $NC_{{\rm tot}}$ attains an explicitly separable form. 

The main problem, one has to deal with at this step is the fact, that in loop quantization there are no quantum counterparts of the connection/triad\footnote{In LQC, due to symmetries, one can introduce a triad operator identifying it with a flux across the unit square. This is however not possible in full LQG.} or the curvature. Thus, before quantizing $NC_{{\rm gr}}$ has to be rewritten in terms of holonomies and fluxes in a process known as the \emph{Thiemann regularization}. Its main drawback is that there is no unique distinguished way of performing it -- it is always based on chosen construction. 

Let us start with the Euclidean part $NH^E$, for which there is one accepted construction choice, discussed in detail in \cite{Ashtekar:2003hd,Ashtekar:2006uz} with subsequent corrections in \cite{Ashtekar:2006wn}. There, the terms involving triads in \ref{a1w} are rewritten as
\begin{equation}
\epsilon_{ijk}\frac{E^{ai} E^{bj}}{\sqrt{det(h)}}=\sum_k \frac{4\sgn(p)}{\kappa\lambda V_0^{\frac{1}{3}}}\epsilon^{abc}\delta^k_c {\rm Tr}\Big( h_i^{(\lambda)}\{(h_i^{(\lambda)})^{-1},V\}\tau_i \Big) \ ,
\label{equ12}
\end{equation}
while the curvature term $F^k_{ab}$ is approximated via holonomies along small square loop (\emph{a plaquet}) $\square_{ij}$
\begin{equation}
F^k_{ab}=-2 \lim_{Ar_{\square}\rightarrow 0}
{\rm Tr}
\Big(
\frac{h^{(\lambda)}_{\square_{ij}}-1}{\lambda^2 V_0^{\frac{2}{3}}}
\Big)
\tau^k\delta^i_a\delta^j_b
\label{equA}
\end{equation}
where $h^{(\lambda)}_{\square_{ij}}$ is holonomy around square $\square_{ij}$. However, realising the classical limit $Ar_{\square}\rightarrow 0 $, while well defined in classical theory, will not be possible in loop quantization due to the lack of continuity of the family of holonomy operators. Therefore, instead of taking that limit, we set the length of the loop side via a phenomenological input from full LQG demanding, that its physical area be twice the smallest non-zero eigenvalue of the area operator from LQG \cite{Ashtekar:2003hd}. That value is exactly the area gap $\Delta$ introduced in \eqref{eq:vb-def}. This sets the fiducial length of the plaquet edge to
\begin{equation}\label{eq:mubar}
  \lambda = \bar{\mu}(v,b) = \sqrt{\Delta}(\alpha|v|)^{-\frac{2}{3}} \ ,
\end{equation}
which gives the final regularized form of $H^E$
\begin{equation}
  H^E=-\frac{2\sgn(v)}{\kappa \gamma \bar{\mu}^3} \sum_{ijk} \epsilon^{ijk}Tr(h_i^{(\bar{\mu})}h_j^{(\bar{\mu})}(h_i^{(\bar{\mu})})^{-1}(h_j^{(\bar{\mu})})^{-1}h_k^{(\bar{\mu})}\{(h_k^{(\bar{\mu})})^{-1},V\}) \ .
\label{euclidean}
\end{equation}

The quantization procedure itself for $H^E$ has been carried out in detailed in \cite{Ashtekar:2006uz,Ashtekar:2006wn}: after implementing \eqref{eq:mubar} and \eqref{euclidean}, reexpressing quantum holonomies and volumes in terms of operators \eqref{intv} 
\begin{subequations}\label{eq:Vh-op}\begin{align}
  \hat{V} &= \alpha|v|\mathbb{I} \ ,
  &
  \hat{h}_k^{(\bar{\mu})} &= \frac{1}{2}(\hat{N}+\hat{N}^{-1})
    - (\hat{N}-\hat{N}^{-1}) i \tau_k \ ,
  \tag{\ref{eq:Vh-op}}
\end{align}\end{subequations}
and choosing symmetric factor ordering we arrive at the final form
\begin{equation}
  \widehat{NH^E} = 12\pi G\gamma^2\left[\sqrt{|v|}\left(\hat{N}^2-\hat{N}^{-2}\right)\sqrt{|v|}\right]^2 \ .
\end{equation}
At this point the reason for a bit involved choice of representing the connections and triad in \eqref{eq:EA-iso} becomes apparent -- the choice is in fact tailored to implemented regularization procedure.

Now, we proceed with implementing the regularization for the Lorentzian part. Here, however, there is no consensus regarding the prescription choice. For the specified (restriction of the) holonomy-flux algebra and the fundamental representation of holonomies, there are in fact three proposals substantially discussed in the literature. We will introduce and briefly discuss them below.

\subsubsection{Mainstream Loop Quantum Cosmology}
\label{sec:mainstream}

The earliest and most popular regularization prescription in LQC is based on an observation that on the classical level $K^i_a$ can be rewritten as $K^i_a=\frac{1}{\gamma}(A^k_a-\Gamma^i_a)$, thus:
\begin{equation}
    E^{ai} E^{bj}K^j_{[a}K^i_{b]}=\frac{1}{2\gamma^2}\epsilon_{ijk}E^{ai} E^{bj}(F^k_{ab}-\Omega^k_{ab}) \ ,
\label{omega}
\end{equation}
with $\Omega^k_{ab}$ being curvature of spin connection $\Gamma^i_a$. Thanks to that, the Lorentzian part is constituted by two terms: one proportional to the Euclidean part and one corresponding to Ricci curvature (vanishing for a flat geometry). But for flat model $\Omega^k_{ab}=0$ thus Lorentzian part is only proportional to the Euclidean part and can be subsumed into it. Such procedure has been implemented for example in the original studies of the isotropic universe dynamics in LQC \cite{Ashtekar:2006uz,Ashtekar:2006wn}. The final result (for flat geometry) then reads
\begin{equation}
  \hat{T} = \frac{1}{2\gamma^2}\hat{H}^E
\end{equation}

It is however worth pointing out, that this procedure is not tied to LQC only. Discussed splitting of the Lorentzian part of the Hamiltonian constraint holds in full GR, thus can be implemented in full LQG, provided an adequate $3d$ Ricci curvature operator can be constructed. Viable proposals for the latter have however been introduced already in \cite{Alesci:2014aza} making the prescription viable in the full theory context.

The procedure of completing the quantization program and extracting the physical prediction for this prescription has been originally carried out in \cite{Ashtekar:2006uz,Ashtekar:2006wn}. Since these steps will be discussed in this paper in greater detail in the context of a prescription not yet fully analyzed, we will skip it for this one, just recalling the main results. The systematic procedure of solving the quantum Hamiltonian constraint (finding its kernel) had led to a picture, where the field $\phi$ could be used as a matter clock and the evolution (with respect to that clock) was generated by the square root of the operator $\hat{\Theta}=-\widehat{NC_{{\rm gr}}}$ -- a non-negative definite second order difference operator, which has been shown to be selfadjoint \cite{Kaminski:2007ew}, thus generating a unique unitary evolution. The evolution picture resulting from it featured a contracting large semiclassical Universe which at Planckian matter energy densities bounced back into an expanding one (preserving semiclassicality). The exact bounce point is identified by a specific (critical) value of matter-energy density $\rho_c\approx 0.41\rho_{{\rm PL}}$.

\subsubsection{Strict Thieman regularization}
\label{sec:Thiemann}

Another possibility is the algorithm proposed originally by Thiemann \cite{Thiemann:2007zz} for full LQG. There, one expresses $K^i_a$ via a Poisson bracket:
\begin{equation}
  K^i_a=\frac{1}{\kappa \gamma^3}\{A^i_a, \{H^E,V\}\}
\label{kia}
\end{equation}
which in cosmological settings (in improved dynamics scheme \cite{Ashtekar:2006wn}) reduce to:
\begin{equation}
K^i_a=-\frac{2}{3\kappa
\gamma^3\bar{\mu}}h_i^{(\bar{\mu})}\{(h_i^{(\bar{\mu})})^{-1}, \{H^E,V\}\} \ ,
\end{equation}
where $h_i^{(\bar{\mu})}$ is a holonomy along an edge of a plaquet defined in sec.~\ref{sec:constraints} (see eq.~\eqref{eq:mubar}). 
This approach was originally applied in the context of LQC in \cite{Yang:2009fp}
while the mathematical properties of the Hamiltonian constraint and the resulting dynamics were analyzed in detail in \cite{Assanioussi:2019iye}. Unlike in mainstream LQC, here the (square of the) evolution generator $\Theta$ is a difference operator of the fourth order. Furthermore, instead of admitting a unique self-adjoint extension, it admits an entire ($U(1)$ labelled) $1$-dimensional family of them, each extension choice corresponding to a particular reflective condition at $v=\infty$. This means that determining the unitary evolution uniquely requires supplementing additional boundary data. The resulting dynamics picture (the same for each extension) significantly altered the mainstream LQC bounce picture: a large semiclassical contracting universe undergoes a bounce at a nit lower critical density in comparizon to the mainstream LQC (about $0.19\rho_{{\rm Pl}}$) but then, instead of reentering classical expansion, it undergoes very rapid inflation as the quantum gravity effects mimic a very large (about $1.03\ell^{-2}_{{\rm Pl}}$) cosmological constant. The unflating universe reaches infinite scale factor for a finite value of the scalar field time, then reflects from it (with the details of reflection governed by the choice of the selfadjoint extension), deflates back, undergoes a second bounce and only after that reenters the expansion epoch accurately described by GR.

\subsubsection{Yang-Ding-Ma regularization}

Alongside with implementation of the strict Thiemann regularization to isotropic cosmology, \cite{Yang:2009fp} yet another prescription has been proposed. It is based on the idea of approximating the extrinsic curvature 1-form via
\begin{equation}
  K^i_a=\frac{1}{\gamma}A^i_a \ ,
\label{KIA}
\end{equation}
Upon implementing it the Lorentzian part of the gravitational Hamiltonian constraint becomes
\begin{equation}
T=\frac{1}{2\kappa\gamma^2}\int_\mathcal{V}d^3x\frac{E^{ai}
E^{bj}}{\sqrt{det(h)}}A^j_{[a}A^i_{b]} \ .
\end{equation}
On the classical level, it is still equivalent to $T$ provided in \eqref{a1w}. 
In the isotropic setting it reduces to
\begin{subequations}\begin{align}
  T &= -\frac{2\sgn(v)}{\kappa^2\gamma^3}\epsilon^{ijk}\Tr\Big(c\tau_ic\tau_j\{c\tau_k,V\}\Big) \ ,
  &
  c(b,v) &= \frac{6\pi\gamma G\hbar}{\alpha^{\frac{2}{3}}} b |v|^{\frac{1}{3}} \ . 
\end{align}\end{subequations}
In order to rewrite it in terms of the holonomies, the following identities have been implemented \cite{Yang:2009fp}
\begin{subequations}
\begin{align}
  c\tau_i &= \lim_{\bar{\mu}\rightarrow
0}\frac{1}{2\bar{\mu}}\Big(h_i^{(\bar{\mu})}-(h_i^{(\bar{\mu})})^{-1}\Big) \ , 
\label{eq:YDM-c_approx} \\
\{c\tau_k,V\} &= h_k^{(\bar{\mu})}\{(h_k^{(\bar{\mu})})^{-1},V\} \ ,
\end{align}
\end{subequations}
where again, instead of taking the limit $\lambda\to 0$, the value $\lambda=\bar{mu}(v,b)$ provided in \eqref{eq:mubar} was taken.
and thus Lorentzian part of the gravitational Hamiltonian constraint becomes
\begin{equation}
T = \frac{\sgn(v)}{2\kappa^2 \gamma^3
\bar{\mu}^3} \sum_{ijk} \epsilon^{ijk}
\Tr\Big(h_i^{(\bar{\mu})}-(h_i^{(\bar{\mu})})^{-1})(h_j^{(\bar{\mu})}-(h_j^{(\bar{\mu})})^{-1})h_k^{(\bar{\mu})}\{(h_k^{(\bar{\mu})})^{-1},V\}\Big)
\label{lorentzian}
\end{equation}

Preliminary studies of the (possible) dynamics resulting from this prescription have been performed in \cite{Yang:2009fp} already. The evolution picture again features a single bounce as a transition between contracting and expanding epoch (both large semiclassical), just happening at a bit different critical energy density -- about $1.73\rho_{{\rm Pl}}$. However, these studies have been performed via the $0$th order effective dynamics -- a heuristic method that already implicitly assumes the existence of a sufficiently large semiclassical sector of the quantum theory. Neither the structure of the physical Hilbert space nor the uniqueness of the evolution has been probed. What is more worrisome, the studies of the numerical Von Neumann stability of the eigenvalue problem of the quantum evolution generator in \cite{Saini:2019tem} indicated, that the existence of sufficiently large physical Hilbert space, let alone the presence of semiclassical sector, is not trivial. For this reason, we will supplement the missing genuine quantum analysis for this prescription in the subsequent sections.

\section{Yang-Ding-Ma regularization in Loop Quantum Cosmology}
\label{sec:mLQC2}

From now on we will focus on the Yang-Ding-Ma regularization recalled above. Having at our disposal the Hamiltonian constraint in the regularized form we can now proceed with constructing its quantum counterpart and verifying the selfadjointness of its gravitational part. The material presented here consisted a part of one of the authors master thesis \cite{Kowalczyk:2021bwr}.


\subsection{Regularization and \texorpdfstring{$\hat{\Theta}$}{Theta } in volume representation}\label{thetaV}

We start by pointing out that because the quantization of the gravitational part and scalar field part has been carried separately we are able to treat kinematical Hilbert space as a tensor product of $\mathcal{H}_{gr}$ and $\mathcal{H}_{\phi}$. 

By assembling together all the terms of the Hamiltonian constraint using (\ref{eq:Hg}, \ref{eq:C_tot}, \ref{euclidean}, \ref{lorentzian}), promoting holonomies and volumes to operators, substituting them with chosen fundamental operators $\hat{V},\hat{N}$  via \eqref{eq:Vh-op} and promoting $(\phi,p_{\phi})$ to operators we can write the total Hamiltonian constraint as
\begin{equation}\label{eq:Ctot_q}
    \hat{C}_{tot}=\mathds{1}_{gr}\otimes(i\hbar\partial_{\phi})^2-\hat{\Theta}\otimes\mathds{1}_{\phi}
\end{equation}
where $\hat{\Theta}$ in symmetric ordering takes the form:
\begin{equation}
\hat{\Theta}=-3 \pi G \hbar^2 \gamma ^2
\sqrt{|\hat{v}|}\Big(\hat{N}^2|\hat{v}|\hat{N}^2-s\hat{N}|\hat{v}|\hat{N}+2(s-1)|\hat{v}|-s\hat{N}^{-1}|\hat{v}|\hat{N}^{-1}+\hat{N}^{-2}|\hat{v}|\hat{N}^{-2}\Big)\sqrt{|\hat{v}|}
\label{thetamlqc2v}
\end{equation}
with $s=4(1+\gamma^2)\gamma^{-2}$. 

Let us focus for a moment on the properties of $\hat{\Theta}$. From \eqref{thetamlqc2v} we immediately see, that it is a difference operator of the $4$th order. Furthermore, (similarly to the other two prescriptions) it naturally splits $\mathcal{H}_{{\rm gr}}$ onto subspaces (preserved by its action) $\mathcal{H}_{\epsilon}$ of restrictions of gravitational states to those supported on the lattices $\mathcal{L}_{\epsilon}=\{ \epsilon+4n;~~ n \in \mathds{Z} \}$, $\epsilon\in [0,4]$. While $\mathcal{H}_{{\rm gr}}$ is noseparable, each $\mathcal{H}_{\epsilon}$ is separable. Provided, that observables are constructed in a way, that they also leave $\mathcal{H}_{\epsilon}$ invariant, the latter become superselection sectors.
One can thus choose one of them and restrict the analysis to it. For further studies, we choose the one corresponding to $\epsilon=0$. 

Furthermore, the operator \eqref{eq:Ctot_q} is symmetric under transformation $v\to -v$ (triad orientation change). Since there is no need for probing the orientation (no fermionic matter) in constructing observables we will preserve that property. Therefore $\mathcal{H}_{\epsilon=0}$ can be further divided onto two superselection sectors: symmetric and antisymmetric. We choose to work with the symmetric one from now on. 

Choices of different sectors (regarding both $\epsilon$ and symmetry/antisymmetry) have been studied in detail in context of the mainstream LQC regularization (see for example \cite{Ashtekar:2006uz,Bentivegna:2008bg,Olmedo:2011zza,MenaMarugan:2011me}). While the mathematical details (like the exact structure of the Hilbert space or the exact form of the point discrete spectrum of operators) could differ, the predicted physics was in all cases the same.

Consider now an eigenvalue problem for $\hat{\Theta}$ on chosen superselection sector. An eigenfunction $e_{\omega}$ corresponding to the eigenvalue $\omega^2$ (only positive part of $\hat{\Theta}$ can contribute to the kernel of \eqref{eq:Ctot_q}) must satisfy a recurrence relation
\begin{equation}
f_{-4}e_{\omega}(v+4)=f_{-2}e_{\omega}(v+2)+(\omega^2-f_0)e_{\omega}(v)-f_{+4}e_{\omega}(v-4)+f_{+2}e_{\omega}(v-2) \ ,
\end{equation}
where
\begin{subequations}
\begin{align}
f_{+4} &= \sqrt{v}\sqrt{v-4}|v-2| \ ,
&
f_{-4} &= \sqrt{v}\sqrt{v+4}|v+2| \ ,
\\
f_{+2} &= s\sqrt{v}\sqrt{v-2}|v-1| \ ,
&
f_{-2} &=s\sqrt{v}\sqrt{v+2}|v+1| \ ,
\\
f_{0} &= 2(s-1)v^2 \ .
\end{align}
\end{subequations}
Since the functions supported at $v=0$ again decouple, the solutions to this recurrence relation are uniquely determined by the free data specified on
$v=2$ and $v=4$, thus the space of solutions has dimension $2$. 
Attempts to find the solutions $e_{\omega}$ numerically have shown, that the problem is unstable. Generically the solutions grow exponentially as it is shown (for particular eigenvalue $\omega=10$) in Fig.\ref{fig:eig-ivp}.

This is an unfortunate difference with respect to the other two prescriptions discussed here -- in volume representation, the eigenfunctions of $\hat{\Theta}$ cannot be numerically evaluated from initial data.


The instability of these solutions was first identified via a VonNeuman stability analysis in \cite{Saini:2019tem} where the large volume behaviour of the equation has been studied. Here we recall the roots of this equation:
\begin{subequations}
\begin{align}
1,~~&~~ 1 \ , \\
\frac{2+\gamma^2-2\sqrt{1+\gamma^2}}{\gamma^2},~~&~~ \frac{2+\gamma^2+2\sqrt{1+\gamma^2}}{\gamma^2} \ ,
\end{align}
\end{subequations}
where the first pair is made of two identical roots equal to one but in the second pair at least one is clearly greater than unity. This gives hope that physical Hilbert space is spanned by some smaller proper subspace of the space of solutions to the initial value problem. Such expectation has been indeed stated in \cite{Saini:2019tem} and subsequently, the second pair of roots has been excluded for consideration of effective dynamic. Here, we will provide verification to this expectation by studying the properties of $\Theta$ in the momentum $b$ representation -- an approach which has proven to be very successful for other two regularizations \cite{Ashtekar:2007em,Assanioussi:2019iye}.

\subsection{Operator \texorpdfstring{$\hat{\Theta}$}{Theta } in \texorpdfstring{$b$}{b } representation}
\label{thetaB}

\indent We begin with defining a type of a Fourier transformation
of a wave function expressed in $v$ variable to the momentum $b$ as
\begin{equation}
  [\mathcal{F}\psi](b)=\sum_{v\in\mathcal{L}_{\epsilon}}\frac{1}{\sqrt{|v|}}\psi(v)e^{i\frac{vb}{2}}
\end{equation}
where (unlike in the other two prescriptions) the domain of $b$ is a circle of unit radius rather than half unit. The functions symmetric in $v$ are transformed into ones satisfying reflection symmetry about $b=\pi$ (as the point antipodal to $b=0$ on the unit circle), that is
\begin{equation}\label{eq:sym_vb}
  \psi(v) = \psi(-v) \quad 
  \Leftrightarrow \quad [\mathcal{F}\psi](b) = [\mathcal{F}\psi](2\pi-b) \ .
\end{equation}
The inverse of the above transformation is
\begin{equation}
  [\mathcal{F}^{-1}\psi](v) 
  = \frac{1}{\pi}\sqrt{|v|}\int_0^{2\pi}db\psi(b)e^{-i\frac{vb}{2}}
  \label{eq:vb-trans-inv}
\end{equation}
Following \eqref{eq:ip} the scalar product in the volume representation takes the form
\begin{equation}\label{eq:ipv}
  \braket{\psi_1|\psi_2}=\sum_{v\in\mathcal{L}_{\epsilon}} \bar{\psi}_1(v)\psi_2(v)
\end{equation}
which in $b$ representation transforms into
\begin{equation}\label{eq:ipb}
  \braket{\psi_1|\psi_2}=\frac{1}{\pi^2}\sum_{v\in\mathcal{L}_{\epsilon}}|v|\int dbdb^{\prime} \bar{\psi}_1(b^{\prime})e^{i\frac{vb^{\prime}}{2}}\psi_2(b)e^{-i\frac{vb}{2}} \ .
\end{equation}
Unfortunately, in the general form, it cannot be compacted to a single integral due to the presence of the absolute value of volume. One can however introduce projections $P^\pm$ onto positive/negative $v$, that is $[P^\pm\psi](v) = \theta(\pm v)\psi(v)$, which for chosen superselection sector $\epsilon=0$ are orthogonal. Then on each subspace, the product simplifies to a local form
\begin{equation}\begin{split}
\braket{\psi_1|\psi_2}_{\pm}&=\pm\frac{1}{\pi^2}\sum_{v\in\mathcal{L}_{\epsilon}}\int
dbdb^{\prime}
\bar{\psi}_1(b^{\prime})v\psi_2(b)e^{-i\frac{v(b-b^{\prime})}{2}}=\pm\frac{1}{\pi}\int
dbdb^{\prime} \bar{\psi}_1(b^{\prime})(-2i\partial_b)\psi_2(b)\delta(b-b')\\
&=\mp\frac{2i}{\pi}\int db \bar{\psi}_1(b)\partial_b\psi_2(b) \ ,
\label{scalarb}
\end{split}\end{equation}
and the whole scalar product can be expressed as
\begin{equation}\label{eq:ip_b}
  \braket{\psi_1|\psi_2} 
  = \braket{P^+\psi_1|P^+\psi_2}_+ + \braket{P^-\psi_1|P^-\psi_2}_- \ .
\end{equation}
At this point it is worth noting, that since we have chosen to work with the symmetric states, the whole information about the wave function is contained in the positive $v$ domain. Thus, in principle, one could restrict ourselves to work with nonnegative and greater than 2 values of $v$ for which the inner product formula \eqref{scalarb} features the minus sign and only the first term in \eqref{eq:ip_b} is relevant. This feature will become useful further in the paper.

Applying the above decomposition, it is straightforward to find the action of volume and shift operator
\begin{subequations}\label{eq:NVb}\begin{align}
  [\hat{V}\psi](b) &= -[2i\alpha\partial_b (P^+-P^-)\psi](b) \ ,
  &
  [\hat{N}\psi](b) 
  &= [\frac{1}{\sqrt{|\hat{v}|}}\exp(\frac{ib}{2})\sqrt{|\hat{v}|}\psi](b) \ ,
\tag{\ref{eq:NVb}}
\end{align}\end{subequations}
which allows to write the operator $\hat{\Theta}$ as a differential operator of the second order
\begin{equation}\begin{split}
  \hat{\Theta} 
  &= 12 \pi G \hbar^2 \gamma ^2 \Big((
  \sin(b)\partial_b(P^+-P^-))^2-\frac{4(1+\gamma^2)}{\gamma^2}(\sin(\frac{b}{2})\partial_b(P^+-P^-))^2\Big) 
  \\
  &=-12 \pi G \hbar^2 \gamma ^2\Big(f(b)\partial_b^2+h(b)\partial_b\Big)
\label{thetab1}
\end{split}\end{equation}
where
\begin{subequations}\label{eq:fh}\begin{align}
  f(b) &= 4\sin^2(\frac{b}{2})\Big(\frac{1+\gamma^2}{\gamma^2}-\cos^2(\frac{b}{2})\Big)
  &
  h(b) &= \sin(b)\Big(\frac{1+\gamma^2}{\gamma^2}-\cos(b)\Big)
  \tag{\ref{eq:fh}}
\end{align}\end{subequations}
{Note, that $f(b)$ is explicitly positive on $b\in(0,2\pi)$, thus the total constraint \eqref{eq:Ctot_q} is (on the domain $(0,2\pi)\times\mathbb{R}$) a hyperbolic partial differential equation. 
Now, we wish to introduce a new variable as a function of $b$ so that the equation for evolution takes the strict Klain-Gordon (K-G) form. 
After application of the chain rule, alongside with condition
\begin{equation} \label{eq:fx-rel}
    \gamma^2f(b)(\frac{dx}{db})^2=1
\end{equation}
$x(b)$ can be integrated out analytically, giving
\begin{equation}\label{eq:x}
  x(b)=-\tanh^{-1}\Big(\frac{\sqrt{2}\cos(\frac{b}{2})}{\sqrt{2+\gamma^2-\gamma^2\cos(b)}}\Big) \ ,
\end{equation}
where a sign is set by the requirement that $\partial_b x(b)>0$ and the free additive constant has been fixed by demanding that $x(b=\pi)=0$
, making $x(b)$ antisymmetric with respect to $b=\pi$. Under this choice, due to \eqref{eq:sym_vb} the symmetry in $v$ will be equivalent to the symmetry in $x$. On the boundaries of the domain in $b$ the new variable reaches the following limits
\begin{subequations}\label{eq:x-lim}\begin{align}
  \lim_{b\rightarrow 0} x(b) &= -\infty \ , 
  &\lim_{b\rightarrow 2\pi} x(b) &= \infty \ ,
  \tag{\ref{eq:x-lim}}
\end{align}\end{subequations}
thus, the set $(0,2\pi)$ is transformed onto the whole real line. Furthermore, $x(b)$ is differentiable on the whole domain.

In the new coordinate the operator $\hat{\Theta}$ now reads
\begin{equation}\label{eq:Theta-x}
\hat{\Theta}=-12 \pi G \hbar^2\partial_x^2 \ ,
\end{equation}
and the scalar product takes the form
\begin{subequations}\label{scalarx}\begin{align}
  \braket{\psi_1|\psi_2} 
  &= \braket{P^+\psi_1|P^+\psi_2}_+ + \braket{P^-\psi_1|P^-\psi_2}_-
  &
  \braket{\psi_1|\psi_2}_{\pm} 
  &= \mp\frac{2i}{\pi}\int^{\infty}_{-\infty} \rd x \bar{\psi}_1(x)\partial_x\psi_2(x)
\tag{\ref{scalarx}}
\end{align}\end{subequations}
Now it is also easy to reexpress in the new coordinate the volume operator: after expressing $\frac{\rd x}{\rd b}$ as a function of $x$ we get
\begin{equation}
  \hat{V}=-i\frac{\alpha}{\sqrt{\gamma^2+1}}f(x)\partial_x(P^+-P^-)
  \label{Vinmlqc2inx}
\end{equation}
where $f(x)=(\gamma^2\tanh^2(x)+1)\cosh(x)$.

This coordinate system and the newly found simple forms of $\hat{\Theta}$ and $\hat{V}$ will be next used in the studies of the selfadjointness of $\hat{\Theta}$ and the dynamics.

\subsection{Deficiency analysis}
\label{sec:theta-defic}

Let us start with verifying the selfadjointness of $\hat{\Theta}$ first. One of very convenient explicit methods of doing so is the analysis of the deficiency subspaces \cite{Reed:1975uy}.Those are defined in the space dual to $\mathcal{H}_{{\rm gr}}$ as spaces of normalizable functions satisfying
\begin{equation}
  \forall \phi\in\mathcal{D} (\phi^{\pm}|\hat{\Theta}\mp i\mathbb{I}|\phi\rangle = 0 \ ,
\end{equation}
where $\mathcal{D}$ is a domain where the action of $\hat{\Theta}$ is well defined. We choose for it the Schwartz space.

Since the dimensionality of the eigenspaces corresponding to pure imaginary eigenvalues does not change when the eigenvalue is rescaled by a positive real factor.  Thus, instead of working with $\hat{\Theta}$ given by \eqref{eq:Theta-x} directly we can define operator $\tilde{\Theta}$ such that
\begin{equation}
\hat{\Theta} = 12\pi G\hbar^2\tilde{\Theta} \ .
\end{equation}
For that operator, the deficiency functions $\psi^{\pm}$ must satisfy the differential equation
\begin{equation}
  \partial_x^2\psi^{\pm}=\pm i\psi^{\pm} \ ,
\label{psiiplusminus}
\end{equation}
{which is straightforward to solve analytically. The solutions are of the form
\begin{subequations}\label{sol}\begin{align}
  \psi^{+} 
  &= c^+\big(\exp(\frac{1+i}{\sqrt{2}}x)+\exp(-\frac{1+i}{\sqrt{2}}x)\big) 
  =: c^+\psi_o \ ,
  &
  \psi^{-} 
  &= c^-\big(\exp(\frac{1-i}{\sqrt{2}}x)+\exp(-\frac{1-i}{\sqrt{2}}x)\big) 
  = c^-\bar{\psi}_o\ ,
  \tag{\ref{sol}}
\end{align}\end{subequations}
where $c^{\pm}\in \mathbb{C}$. 

In order to probe their normalizability we estimate their behaviour in $v$-representation for large $|v|$. Because of the symmetry of $\psi_{\pm}$ and the form of the solutions \eqref{sol} it is enough to check the function $\psi_o$ for large positive $v$. We proceed with plugging it into the transform \eqref{eq:vb-trans-inv} and using the stationary phase method. First, let us decompose $\psi_o$ onto the modulus and phase
\begin{equation}
  \psi_o(v) 
  = \frac{\sqrt{|v|}}{\pi}\int_0^{2\pi} \rd b |\psi_o(b)|e^{i\varphi(b)} e^{-\frac{vb}{2}} \ .
\end{equation}
The dominant contributions to the above integral will come from the neighbourhoods of the critical points, where $\partial_b\varphi(b) = v/2$. Due to the form of $\psi_o$ those correspond to large positive value of $x(b)$, for which $\phi_o(b)$ can be very accurately approximated by its first (exponentially amplified) term. Thus for large $v$
\begin{equation}\label{eq:psi_o-approx}
  \psi_o(v) \approx \int_0^{2\pi} \rd b \, e^{\frac{x(b)}{\sqrt{2}}} e^{i\left(\frac{x(b)}{\sqrt{2}}-\frac{vb}{2}\right)}  \ .
\end{equation}
The critical points (for which $[\partial_bx](b_\star)-v/2=0$) are the solutions to the equation
\begin{equation}
  f(b) = \frac{4}{\gamma^2v^2} \ ,
\end{equation}
where $f$ is a function defined in \eqref{eq:fh}. For large $v$ there are two solutions: one close to $b=0$ and one close to $b=2\pi$, however, we can neglect the first one as near it the integrand of \eqref{eq:psi_o-approx} will be exponentially suppressed. 

Expanding function $f(b)$ (up to a leading term) we find the approximate solution for the critical point of interest 
\begin{equation}\label{eq:bstar}
  b_\star \approx 2\pi - \frac{2}{v} \ .
\end{equation}
Similarly expanding the trigonometric functions in \eqref{eq:x} and rewriting artanh in terms of logarithms we arrive to an approximation on $x(b_\star)$
\begin{equation} \label{eq:xbstar}
  x(b_\star) \approx \frac{1}{2}\ln(v^2) \ .
\end{equation}
Using the relation \eqref{eq:fx-rel}, the formula \eqref{eq:fh} for $f(b)$, applying the expansion into power series and substituting \eqref{eq:bstar} we also obtain estimates of the derivatives of $x$
\begin{subequations}\label{eq:xdbstar}\begin{align} 
  x'(b_\star) &\approx \frac{v}{2} \ ,
  &
  x''(b_\star) &\approx \frac{v^2}{4} \ .
  \tag{\ref{eq:xdbstar}}
\end{align}\end{subequations}
Now, applying the stationary phase approximation to \eqref{eq:psi_o-approx} 
for the critical point $b_\star$ we get
\begin{equation}
  \psi_o(v) \approx \frac{\sqrt{v}}{\pi} e^{\frac{x(b_\star)}{\sqrt{2}}}
  \sqrt{\frac{2\sqrt{2}\pi}{x''(b_\star)}}
  \exp\left[i\left(\frac{x(b_\star)}{\sqrt{2}}-\frac{vb_\star}{2}+\frac{\pi}{4}\right)\right] \ ,
\end{equation}
and finally substituting \eqref{eq:xbstar} and \eqref{eq:xdbstar} we arrive to an estimate on the magnitude for large positive $v$
\begin{equation}
  |\psi_o(v)| \sim v^{\frac{\sqrt{2}-1}{2}} \ . 
\end{equation}
Since the power is strictly positive, $\psi_o$ is not normalizable on $\mathcal{H}_{{\rm gr}}$. 
Thus, $\braket{\psi^{+}|\psi^{+}}$ and $\braket{\psi^-|\psi^{-}}$ are not finite unless both the coefficients $c^{\pm}$ vanish.
Thus, both deficiency subspaces are spaned only by a zeroth vector, meaning
\begin{subequations}\label{eq:dimK}\begin{align}
  \dim(\mathcal{K}_+) &= 0 \ ,
  &
  \dim(\mathcal{K}_{-}) &= 0 \ .
  \tag{\ref{eq:dimK}}
\end{align}\end{subequations}
As a consequence, the operator $\hat{\Theta}$ is essentially selfadjoint on the domain $\mathcal{D}$ \cite{Reed:1975uy}.

Having established the selfadjointness of $\hat{\Theta}$ we can now calculate its spectrum and calculate the basis of $\mathcal{H}_{{\rm gr}}$ following from its spectral decomposition.

\subsection{Energy basis on \texorpdfstring{$\mathcal{H}_{{\rm gr}}$}{Gravitational Hilbert Space }}
\label{sec:eigenbasis}

For the reasons, that will become apparent in sec.~\ref{sec:Hphys} (essentially due to negative definiteness of $\partial_{\phi}^2$ and the form of the Hamiltonian constraint \eqref{eq:Ctot_q}) only the positive part of the operator $\Theta$ is relevant for the description of the studied system. As a consequence, the eigenvalue problem relevant for its spectral decomposition can be formulated as
\begin{equation}
  p_{\phi}^2\psi = \omega^2\hbar^2\psi = \hat{\Theta}\psi 
\end{equation}
Given the form \eqref{eq:Theta-x} of $\Theta$ solving it in $b$-representation is straightforward
\begin{equation}
  \psi_k(b) = N^+_k e^{ikx(b)} + N^-_k e^{-ikx(b)} , \quad \omega(k) = \sqrt{12\pi G} k , \quad k\geq 0 \ ,
\end{equation}
where $N^{\pm}_k$ are normalization constants. In particular, the symmetric eigenstates will be described by the following eigenfunctions
\begin{equation}\label{eq:eig-sym-x}
  \psi_k(b) = N_k \cos(kx(b)) \ ,
\end{equation}
where $N_k$ is again a normalization constant, thus in the symmetric sector the eigenspaces are nondegenerate.

As in $b$-representation the inner product on $\Hil_{\rm gr}$ does not have a simple local form, normalizing \eqref{eq:eig-sym-x} is a bit involved. 
First, by transforming $\psi_k$ back to $v$ representation (see Appendix~\ref{sec:WDW-limit}) we observe, that in large $v$ limit it converges to certain combination (reflected plane wave) of eigenfunctions of the geometrodynamical (Wheeler-DeWitt) analog of studied model (recalled in more detail in Appendix~\ref{sec:WDW}), namely from \eqref{eq:e-LQC-as} we have 
\begin{equation}\label{eq:psik-v}\begin{split} 
  \psi_k(v) 
  &= N_k \frac{4}{\sqrt{\pi}} k \sinh\left(\frac{k\pi}{2}\right) 
    \left[ \Gamma(-ik) e^{-ik(\sigma_0+\ln(2))} \ub{e}_k(v) + \Gamma(ik) e^{ik(\sigma_0+\ln(2))} \ub{e}_{-k}(v) \right] 
    + O(|v|^{-3/2})  \\
  &= N_k \left(2\sqrt{k} + O(e^{-\frac{k\pi}{2}})\right) 
  \left[ e^{-ik\phi_o(k)} \ub{e}_k(v) + e^{ik\phi_o(k)} \ub{e}_{-k}(v) + O(|v|^{-3/2}) \right] \ ,
\end{split}\end{equation}
where $\ub{e}_k$ are the elements of the orthonormal basis of gravitational Hilbert space in the WDW analog of the model \eqref{eq:WDW-basis} and the $k$-dependent phase offset $\phi_o$ is defined via relation
\begin{equation}\label{eq:Gamma-dec}
  \Gamma(ik) e^{-ik(\sigma_0+\ln(2))} =: |\Gamma(ik)| e^{-ik\phi_o(k)} = \sqrt{\frac{\pi}{k\sinh(k\pi)}} e^{-ik\phi_o(k)} \ ,
\end{equation}
while $\sigma_0$ is given by \eqref{eq:sigma0-def}. 

By inserting the above formula into the inner product on $\Hil_{{\rm gr}}$, noting that the contributions involving the remnant are fine and approximating the sum over (portion of the) lattice $\mathcal{L}_{\epsilon=0}$ by an integral we arrive to the estimate \eqref{eq:ip-app}
\begin{equation}\begin{split}
  \langle\psi_k|\psi_{k'}\rangle 
  &= \frac{16kN_k^2}{\sinh(k\pi)} \sinh^2\left(\frac{k\pi}{2}\right) \delta(k-k') + f(k,k') \\
  &= 8kN_k^2(1+O(e^{-k\pi/2})) \delta(k-k') + f(k,k') \ ,
\end{split}\end{equation}
where $f$ is a function. 
This, the asymptotic behaviour \eqref{eq:psik-v} and the fact that the spectrum of the WDW counterpart $\ub{\Theta}$ of $\Theta$ (see Appendix~\ref{sec:WDW}) is continuous and ${\rm Sp}(\ub{\Theta})=\re^+$ allows us to conclude, that the (positive part of the) spectrum of $\Theta$ is also continuous and ${\rm Sp}(\Theta) = \re^+$. In consequence the function $f$ must vanish and we can explicitly construct the orthonormal basis ${e_k}_{k\in\re^+}$, where \eqref{eq:LQC-normalization}
\begin{equation}\label{eq:LQC-basis}
  e_k(b) = \frac{1}{2\sqrt{k}} \frac{\sqrt{\sinh(k\pi)}}{\sinh(k\pi/2)} \cos(kx(b)) 
  = \frac{1}{\sqrt{2k}} \left( 1 + O(e^{-k\pi/2}) \right) \cos(kx(b)) \ .
\end{equation}

Having at our disposal elements of the basis in gravitational Hilbert Space we can now make a final comment on the instability of the eigenvalue problem described in subsection \ref{thetaV}. By choosing to work with symmetric eigenstates described by formula \ref{eq:eig-sym-x} we can go back to volume representation by applying transform \ref{eq:e-inv-app}. A result of such transformation (the basis element $e_k(v)$ for $k=20$) is presented in Fig.~\ref{fig:eig-brep} (marked by red dots) alongside with combination of  eigenfunctions of the Wheeler-DeWitt model (blue line) described via \eqref{eq:e-LQC-as} which in fact is a large $v$ limit to which chosen basis element converges. One can in particular see the same qualitative features as those of evolution operator eigenfunction in mainstream LQC: $(ii)$ a suppression below the bounce point, $(ii)$ regular oscillatory behavior in classically allowed region, and $(iii)$ the convergence to a certain standing wave in the Wheeler-DeWitt analog of the model studied. It is noteworthy, that for this prescription standard numerical methods allow to reliably determine the eigenfunctions only for relatively small $k$. For larger ones the numerical noise partially masks the sub-bounce suppression, not allowing for reliable application of the result in the genuine quantum analysis like the one of \cite{Ashtekar:2006uz}. 
Nevertheless, (as the analytical studies of sec.~\ref{sec:eigenbasis} have shown) this function behaves well on it's respective domain and whole set with a scalar product will form a proper Gravitational Hilbert space which is subspace of solutions to \eqref{eq:Ctot_q}. This deficiency of numerical methodology is the main reason, why the analytical method of probing the dynamics has been used instead.
\begin{figure}[H]
  \begin{centering}
    \subfloat[]{\includegraphics[width=.45\textwidth]{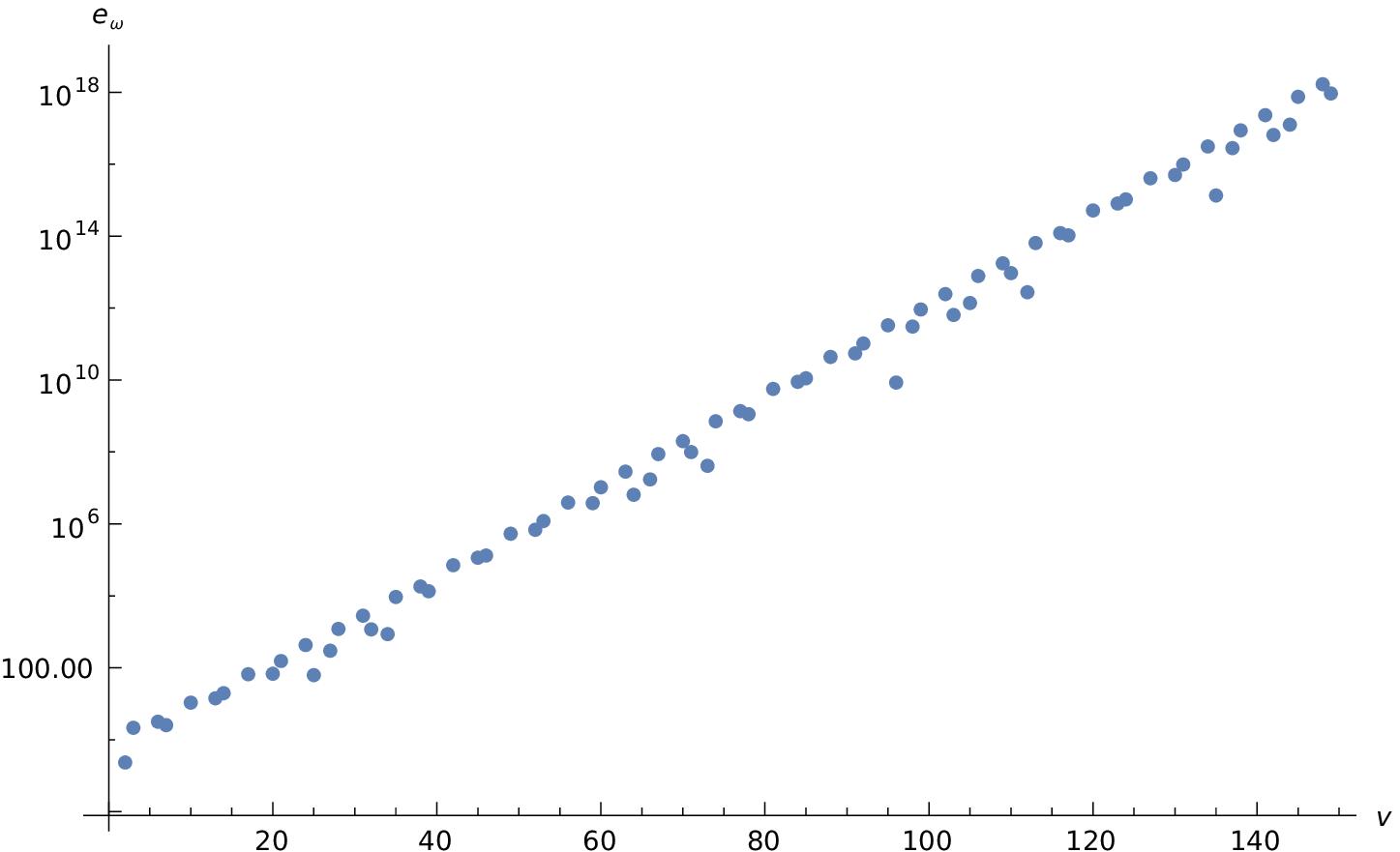}\label{fig:eig-ivp}}
    \hspace{0.5cm}
    \subfloat[]{\includegraphics[width=0.45\textwidth]{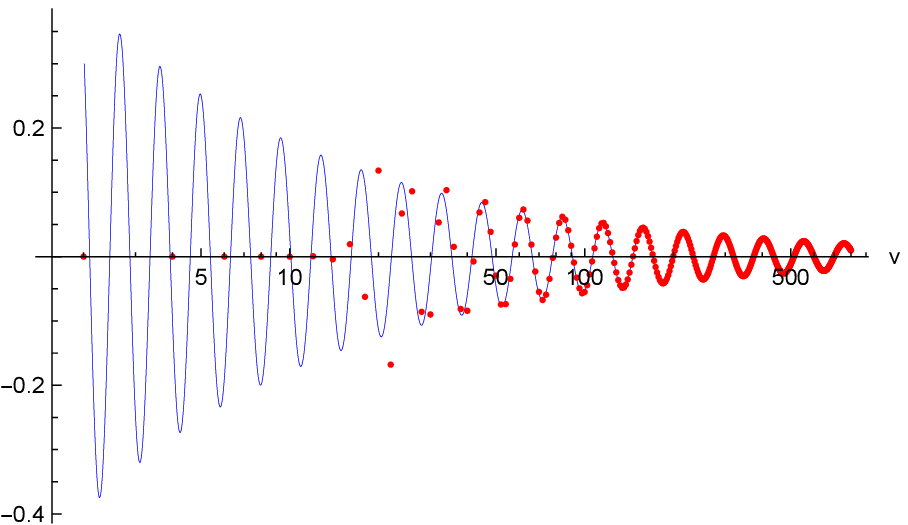}\label{fig:eig-brep}}
  \end{centering}
  \label{figa4}
  \caption{Plot of an example of a (generic) solution to an initial value problem corresponding to eigenvalue equation (\ref{fig:eig-ivp}, here corresponding to $\omega=10$) is compared against the actual element of the energy eigenbasis (\ref{fig:eig-brep}, an example of $e_k(v)$ for $k=20$). In the latter we observe the exponential suppression in classically forbidden (sub-bounce) region and a convergence to a WDW standing wave in large volume limit.}
\end{figure}

Since many relevant physical observables (for example volume) involve the projection operators $p^{\pm}$ onto a particular orientation of $v$, it is necessary to determine, how these projections act on our basis elements. For that, let us introduce a pair of projectors $\tilde{P}^{\pm}$ projecting onto the subspaces of lefthand/righthand moving plane waves in $x$ coordinate. On the symmetric basis element, they act as follows
\begin{equation}\label{eq:tP-proj}
  [\tilde{P}^{\pm} e_k](b) = \frac{N_k}{2} e^{\pm ikx(b)} \ ,
\end{equation}
where $N_k$ is again provided by \eqref{eq:LQC-normalization}. By relatively simple calculations (see again Appendix~\ref{sec:WDW-limit}) one can show, that (following \eqref{eq:proj-brep})
\begin{equation}\label{eq:proj-brep-main} 
  [P^{\pm} e_k](b) = \frac{\sinh(k\pi/2)}{\sinh(k\pi)} 
  \left[ e^{\pm\frac{k\pi}{2}} [\tilde{P}^+e_k](b) + e^{\mp\frac{k\pi}{2}} [\tilde{P}^-e_k](b) \right] 
  = [\tilde{P}^{\pm}e_k](b) + O(\frac{e^{-k\pi}}{\sqrt{k}}) 
  \ .
\end{equation}
In consequence, the action of $P^{\pm}$ on states peaked about large $k$ (the class relevant for the description of large universes) is with great accuracy approximated by the action of $\tilde{P}^{\pm}$.

\section{Dynamics}
\label{Dynamics}

Having at our disposal the quantum Hamiltonian constraint \eqref{eq:Ctot_q} of which all components are essentially selfadjoint, we now can proceed with the next step of the Dirac program -- solving the quantum constraint and constructing a set of physically meaningful observables. For this, we can directly apply the procedure already used in mainstream LQC \cite{Ashtekar:2006uz} originating from group averaging techniques \cite{Ashtekar:1995zh}.

\subsection{Physical Hilbert space and observables}
\label{sec:Hphys}

Given that the quantum Hamiltonian constraint has an explicit separable form \eqref{eq:Ctot_q} where both component operators are essentially selfadjoint, it is also essentially selfadjoint, thus its exponents $U(\lambda):=exp(i\lambda\hat{C}_{{\rm tot}}$ form a group of unitary transformations. Then one can build a rigging map \cite{Ashtekar:1995zh}
\begin{equation}
  \mathcal{H}_{{\rm phy}}^* \ni \eta|\psi\rangle 
  = \left(\int_{\mathbb{R}}\rd\lambda U(\lambda)|\psi\rangle\right)^\dagger
  = \int_{\mathbb{R}}\rd\lambda\langle\psi| e^{-i\lambda\hat{C}_{{\rm tot}}} \ ,
\end{equation}
projecting onto the (dual of the) physical Hilbert space, of which inner product is also induced via $\eta$ 
\begin{equation}\label{eq:ga-ip}
  (\eta\psi|\eta\chi) 
  = \int_{\mathbb{R}}\rd\lambda \langle\chi|U^{-1}(\lambda)|\psi\rangle
  = \int_{\mathbb{R}}\rd\lambda \langle\chi|e^{-i\lambda\hat{C}_{{\rm tot}}}|\psi\rangle
\end{equation}

On the Hilbert space $\Hil_{\rm gr}$ we choose an orthonormal basis of the eigenstates $(e_k|$ of the $\Theta$ operator (given in \eqref{eq:LQC-basis}) and the eigenstates $(f_\omega|$ of the operator $i\partial_{\phi}$ (corresponding to the eigenvalue $\omega$) and decompose the states $|\psi\rangle$, $|\chi\rangle$ in it
\begin{equation}
  ((e_k|\otimes(e_{\omega}|)|\psi\rangle =: \psi(k,\omega) \ , \quad 
  ((e_k|\otimes(e_{\omega}|)|\chi\rangle =: \chi(k,\omega) \ , \quad
  k\in\re^+,\ \omega\in\re \ .
\end{equation}
The action of the constraint $\hat{C}_{\rm tot}$ on the chosen basis yields
\begin{equation}
  ((e_k|\otimes(e_{\omega}|)\hat{C}_{\rm tot} = \hbar^2(\omega^2-12\pi Gk^2)((e_k|\otimes(e_{\omega}|)
\end{equation}
which in turn (together with the orthonormality of selected basis) allows to easily evaluate the integral \eqref{eq:ga-ip}
\begin{equation}
  (\eta\psi|\eta\chi) 
  = \int_{\re^+} \frac{\rd k}{k} \bar{\psi}(k,\beta k) \chi(k,\beta k) 
  + \int_{\re^+} \frac{\rd k}{k} \bar{\psi}(k,-\beta k) \chi(k,-\beta k) \ , 
\end{equation}
where $\beta = \sqrt{12\pi G}$. As a consequence, each physical state can be represented by a pair of spectral profiles. For example in $(v,\phi)$ representation   
\begin{equation}\label{eq:phys-gen}
  \psi(v,\phi) = \int_{\re^+} \rd k \psi^+(k) e_k(v) e^{i\beta k\phi}
  + \int_{\re^+} \rd k \psi^-(k) \bar{e}_k(v) e^{-i\beta k\phi} \ .
\end{equation}
Per similarity with the Klain-Gordon equation (with $\phi$ being an analog of time), the two above terms are interpreted as positive/negative energy ones respectively and constitute superselection sectors. As a consequence one usually restricts attention to the positive energy sector only. We will apply this restriction in further analysis (thus setting $\psi^-(k)=0$ and dropping the $+$ marker in $\psi^+(k)$), though the negative sector can be easily included with just minor adjustments to the calculations.\footnote{In the case when the evolution is parametrized by a massless scalar field restriction to positive energy sector only may not admit semiclassical sector \cite{Kaminski:2019tqo}. In such cases adding small tails in the negative energy sector is needed in order to regularize the infinities and restore the large subset of semiclassical states. This issue will be discussed in more details further in the paper.}

The constraint $\hat{C}_{\rm tot}$ can be rewritten as a product of commuting operators
\begin{equation}
  \hat{C}_{\rm tot} = \hat{C}^+_{\rm tot} \hat{C}^-_{\rm tot} \ , \qquad 
  \hat{C}^{\pm}_{\rm tot} := \pm i\hbar \mathbb{I} \otimes \partial_{\phi} - \sqrt{\Theta} \otimes \mathbb{I} \ ,
\end{equation}
where $\hat{C}^{\pm}_{\rm tot}$ annihilates the positive/negative energy parts of the physical state respectively. Thus the restriction to the positive energy sector can be implemented already on the level of group averaging procedure by selecting the rigging map
\begin{equation}
  \mathcal{H}_{{\rm phy}}^* \ni \eta^+|\psi\rangle 
  = \left(\int_{\mathbb{R}}\rd\lambda U^+(\lambda)|\psi\rangle\right)^\dagger
  = \int_{\mathbb{R}}\rd\lambda\langle\psi| e^{-i\lambda\hat{C}^+_{{\rm tot}}} \ ,
\end{equation}
which yields the space of physical states corresponding to \eqref{eq:phys-gen} with $\psi^-$ set to zero and the physical scalar product 
\begin{equation}\label{scalark}
  \langle\eta\psi|\eta\chi\rangle = \int_{\re^+} \rd k \bar{\psi}(k)\chi(k) \ ,
\end{equation}
which is the textbook inner product in the Klein-Gordon system. 

The structure above is the same as the decomposition with respect to the energy basis of states in $\Hil_{\rm gr}$, thus one can easily construct a unitary (inner product preserving) transformation
\begin{equation}
  U_{\phi} : \Hil_{\rm phy} \to \Hil_{\rm gr} \ , \quad [U_{\phi}\Psi](k) = \Psi(k)e^{i\beta k\phi} \ ,
\end{equation}
casting the wave function of the physical state onto its constant $\phi$ slice. Thus, the physical state can be reinterpreted as an evolution (with respect to the field $\phi$) of geometry states. That evolution is described by a family of unitary operators $U_{\phi_o,\phi}:=U^{-1}_{\phi}U_{\phi_o}$. Such procedure is known as a \emph{deparametrization on a quantum level} and the field $\phi$ attains a role of an \emph{internal clock}. 

This picture further allows us to easily construct out of any geometry observable $\hat{O}$ a family of Dirac observables $\hat{O}_{\phi} = U_{\phi}^{-1}\hat{O}U_{\phi}$ measuring ``quantity $O$ at a given moment of $\phi$''\footnote{Such observables can also be constructed directly through group averaging, see for example \cite{Kaminski:2009qb} for details} -- the so-called Partial observables \cite{Rovelli:2001bz,Dittrich:2004cb}. Following the existing LQC literature, we will focus our attention on the volume at fixed $\phi$ and the scalar field momentum (measuring the ``energy'' with respect to the scalar field ``time'') of whose action on the state represented by a wave function in $(x,\phi)$ variables reads
\begin{equation}
  \hat{V}_{\phi_0}\Psi(x,\phi) := U^{-1}_{\phi_0} \hat{V} U_{\phi_0}\Psi(x,\phi) = \alpha e^{i(\phi-\phi_0)\sqrt{\Theta}}|v|\Psi(x,\phi_0) \ ,
  \qquad
  \hat{p}_{\phi}\Psi(x,\phi) = -i\hbar\partial_{\phi}\Psi(x,\phi)
  \label{Diracsmlqc2}
\end{equation}
where $\alpha$ is the constant defined in \eqref{eq:vb-def}.
Note, that on physical states $\hat{p}_{\phi} = \sqrt{\Theta}$, thus $\hat{p}_{\phi}$ is a constant of motion.

In order to probe the system dynamics we need to evaluate the expectation values of the selected observables \eqref{Diracsmlqc2}. In practice, evaluating them  boils down to evaluation of the expectation values of the kinematical observables $\hat{V}$, $\sqrt{\Theta}$ at the respective slices of constant value of $\phi$. Since the states most relevant from the cosmological point of view are those that are semiclassical in at least one epoch of the evolution and describe large universe. Thus in further studies, we will focus on the states that are sharply peaked at $p_{\phi}\gg \sqrt{G}\hbar$ and for which at least at one $\phi$ $\langle \hat{V}_{\phi}^n \rangle < \infty$. As the wave function has a simple analytic form in variable $x$ we will be performing the calculations either in $x$-representation or directly in the energy spectrum of $\Theta$ (further denoted as $k$-representation).

\subsection{Observables in energy representation}
\label{sec:obs-k}

In $x$ variable the volume operator takes (modulo the projections $P^\pm$) a quite simple form \eqref{Vinmlqc2inx}. Furthermore, by \eqref{eq:proj-brep-main} the projections $P^\pm$ can be approximated by the projections onto lefthand/righthand moving plane waves $\tilde{P}^{\pm}$ of which action on the chosen basis is again simple \eqref{eq:tP-proj}. This in particular allows to write an approximation of the inner product as
\begin{equation}\label{eq:ip-approx}\begin{split}
  \langle\Psi_1|\Psi_2\rangle &= \langle U_{\phi}\Psi_1|U_{\phi}\Psi_2 \rangle 
  =: \langle \Psi_{1\phi}|\Phi_{2\phi}\rangle  \\
  &= \frac{2i}{\pi}\left[ \int_{-\infty}^{\infty} \rd x [\tilde{P}^-\Psi_{1\phi}](x)\partial_x[\tilde{P}^-\Psi_{2\phi}](x) 
  - \int_{-\infty}^{\infty} \rd x [\tilde{P}^+\Psi_{1\phi}](x)\partial_x[\tilde{P}^+\Psi_{2\phi}](x) \right] 
  \left(1 + O(e^{-\langle k\rangle \pi}) \right) \\ 
  &= -\frac{4i}{\pi} 
  \left[ \int_{-\infty}^{\infty} \rd x [\tilde{P}^+\Psi_{1\phi}](x)\partial_x[\tilde{P}^+\Psi_{2\phi}](x) \right] 
  \left(1 + O(e^{-\langle k\rangle \pi}) \right) \ ,
\end{split}\end{equation}
where in the 1st equality we applied the (straightforward to see in $v$-representation) equality between physical \eqref{scalark} and geometry inner product and in the 3rd one we approximated the form \eqref{scalarx} of the geometry inner product with $P^{\pm}$ substituted with $\tilde{P}^{\pm}$. In the last equality, we took advantage of the symmetry of considered states. 

For further simplification let us consider an approximation of the operators $\tilde{P}^{\pm}$ where $N_k$ in \eqref{eq:tP-proj} is replaced with its approximation, that is
\begin{equation}\label{eq:proj-final}
  [\mathcal{P}^{\pm} e_k] (x) := \frac{1}{2\sqrt{2k}} e^{\pm ikx} \quad \Rightarrow \quad 
  [\mathcal{P}^{\pm}\psi](x) = \frac{1}{2\sqrt{2}}\int_0^{+\infty} \rd k \frac{\psi(k)}{\sqrt{k}} e^{\pm ikx} 
\end{equation}
Then, we can introduce an approximate inner product 
\begin{equation}\label{scalarxapproximate}
  \langle\Psi_{1\phi}|\Psi_{2\phi}\rangle^{\sim} :=
  -\frac{4i}{\pi} 
  \int_{-\infty}^{\infty} \rd x [\mathcal{P}^+\Psi_{1\phi}](x)\partial_x[\mathcal{P}^+\Psi_{2\phi}](x)
\end{equation}
which approximates the exact inner product \eqref{eq:ip-approx} again up to the (multiplicative) 
correction $(1+O(e^{-\langle k\rangle \pi}))$.

The operator $\hat{V}$ can be approximated in a similar fashion (with the correction of the same order). Before doing that, let us however consider its slightly more general form, namely
\begin{equation}
    \hat{V}=-iAf(x)\partial_x(P^+-P^-) \ ,
    \label{eq:V-gen}
\end{equation}
where $A$ and $f(x)$ are respectively constant and function depending on $x$. An advantage of such generalization is that we will be able to apply the results of the calculations to all the prescriptions presented in this article, just inserting the value of $A$ and the form of $f(x)$ appropriate for the prescription we chose to analyze. Then, we approximate \eqref{eq:V-gen} by
\begin{equation}
  \tilde{V} = -iAf(x)\partial_x(\mathcal{P}^+-\mathcal{P}^-) \ .
\end{equation}
Note, that when evaluating the expectation values $\langle \hat{V}^n \rangle^{\sim}$ on the symmetric states we can again approximate the operator $\tilde{V}$ by 
\begin{equation}\label{generalformofvolumoperator}
  \tilde{V}^+ := -iAf(x)\partial_x\mathcal{P}^+ \ ,
\end{equation}
as the corrections due to neglecting crossterms $\mathcal{P}^-\mathcal{P}^+$ are of the same order as 
in the previous approximations. Thus for the calculations of the quantum trajectories $V(\phi) = \langle\Psi\hat{V}_{\phi}\Psi\rangle$ we will be using that form of the operator. 

For further evaluations, it is more convenient to rewrite the expectation value of $\tilde{V}^\sim$ in explicitly symmetric form
\begin{equation}
  \braket{\tilde{V}^+} = -\frac{4A}{\pi}\int^{\infty}_{-\infty} \rd x
  [\mathcal{P}^+\bar{\Psi}](x)\partial_x \Big(f(x)\partial_x[\mathcal{P}^+\Psi](x)\Big)
  = \frac{4A}{\pi}\int^{\infty}_{-\infty} \rd x 
  (\partial_x[\mathcal{P}^+]\bar{\Psi}](x)) f(x)\partial_x[\mathcal{P}^+\Psi](x) \ .
\label{vgeneral1}
\end{equation}
Plugging the form \eqref{eq:proj-final} of $\mathcal{P}\psi$ to the above we get
\begin{equation}\label{eq:Vtraj1}\begin{split}
  \braket{\tilde{V}^+_{\phi}} 
  = \braket{U_{\phi}\Psi|\tilde{V}^+U_{\phi}\Psi} 
  &= \frac{A}{2\pi}\int^{\infty}_{-\infty}\int_{0}^{\infty}\int_{0}^{\infty}
  \rd x\rd k\rd k'
  \Big(\partial_x\frac{\bar{\Psi}(k')}{\sqrt{k'}}e^{-ik'x}e^{-i\beta k'\phi}\Big)
  f(x) \partial_x\frac{\Psi(k)}{\sqrt{k}}e^{ikx}e^{i\beta k\phi} \\
  &=\frac{A}{2\pi}\int^{\infty}_{-\infty}\int_{0}^{\infty}\int_{0}^{\infty}
  \rd x\rd k\rd k'
  \bar{\Psi}(k'){\sqrt{k'}}f(x)\Psi(k){\sqrt{k}}e^{i(k-k')(x+\beta\phi)} \ .
\end{split}\end{equation} 
Our goal is to rexpress the trajectory as a function of $\phi$ and come constants of motion. One way of achieving that is to rewrite the above expression (including all the operators) entirely in $k$-representation. This requires performing an integral over $x$ explicitly. In order to do that we 
expand function $f$ into Taylor series 
\begin{equation}
  f(x)=\sum_{n=0}^{\infty}\frac{1}{n!}f^{(n)}(0)x^n \ ,
\end{equation} 
and further note, that the action of $\hat{x}\mathcal{P}^+U_{\phi}$ can be written as
\begin{equation}\begin{split}
  [\hat{x}\mathcal{P}^+U_{\phi}\Psi](x,\phi)
  &= \frac{1}{2\sqrt{2}} \int_{0}^{\infty} \rd k x\frac{\psi(k)}{\sqrt{k}}e^{ik(x+\beta\phi)}
  = \frac{1}{2\sqrt{2}} \int_{0}^{\infty} \rd k \frac{\psi(k)}{\sqrt{k}}(-i\partial_k-\beta\phi)e^{ik(x+\beta\phi)} \\
  &= [(i\partial_k-\frac{i}{2k}-\beta\phi\mathbb{I})\mathcal{P}^+U_{\phi}\Psi](x,\phi) \\ 
\end{split}\end{equation}
Analogously 
\begin{equation}
  [\hat{x}^n\mathcal{P}^+U_{\phi}\Psi](x,\phi)
  = \frac{1}{2\sqrt{2}} \int_{0}^{\infty} \rd k 
  \frac{\psi(k)}{\sqrt{k}}(-i\partial_k-\beta\phi)^n  e^{ik(x+\beta\phi)} \ .
\end{equation}
Plugging the above into \eqref{eq:Vtraj1} we get
\begin{equation}\label{expectationV}\begin{split}
  \braket{\tilde{V}^+_{\phi}}
  &= \frac{A}{2\pi}\sum_{n=0}^{\infty}\int^{\infty}_{-\infty}\int_{0}^{\infty}\int_{0}^{\infty}
  \rd x \rd k \rd k'
  \bar{\psi}(k'){\sqrt{k'}}e^{-ik'x}e^{-i\beta k'\phi}\frac{(-1)^n}{n!}f^{(n)}(0)\psi(k){\sqrt{k}}(i\partial_k+\beta\phi)^ne^{ikx}e^{i\beta k\phi}\\
  &= \frac{A}{2\pi}\sum_{n=0}^{\infty}\int^{\infty}_{-\infty}\int_{0}^{\infty}\int_{0}^{\infty}
  \rd x \rd k \rd k'
  \bar{\psi}(k'){\sqrt{k'}}\frac{1}{n!}f^{(n)}(0)(i\partial_k+\beta\phi)^n\psi(k){\sqrt{k}}e^{i(k-k')x}e^{i\beta (k-k')\phi}\\
  &= A \sum_{n=0}^{\infty}\sum_{l=0}^{n}\binom{n}{1}
  \frac{1}{n!}f^{(n)}(0)(\beta\phi)^{n-l}\int_{0}^{\infty}
  \rd k \bar{\psi}(k){\sqrt{k}}(i\partial_k)^l\psi(k){\sqrt{k}}\\
  &= A \sum_{l=0}^{\infty}\frac{1}{l!}f^{(l)}(\beta\phi)\int_{0}^{\infty}
  \rd k \bar{\psi}(k){\sqrt{k}}(i\partial_k)^l{\sqrt{k}}\psi(k)\\
  &=A\sum_{l=0}^{\infty}\frac{1}{l!}f^{(l)}(\beta\phi)\braket{{\sqrt{k}}(i\partial_k)^l{\sqrt{k}}} \ ,
\end{split}\end{equation}
where to get to the expressions in the 2nd, 3rd and 4th line we respectively: performed integration by parts $n$ times (in order to act on the spectral profile), performed a binomial decomposition of $(i\partial_k+\beta\phi)^n$ and finally regrouped and summed up all the terms by the order of the derivative of $f$. In the last one, we identified the integral as the expectation value of an operator ${\sqrt{k}}(i\partial_k)^l{\sqrt{k}}$.  

Similar, though a bit more tedious calculations can be performed in order to find the trajectory of $(\tilde{V}^+_{\phi})^2$, needed to determine the dispersion (variance) of the state. The repetition of all the steps performed for \eqref{expectationV} yields
\begin{equation}
  \braket{\tilde{V^+_{\phi}}^2}=\frac{1}{2}A^2\sum_{l=0}^{\infty}\frac{1}{l!}h^{(l)}(\beta\phi)\braket{{\sqrt{k^3}}(i\partial_k)^l{\sqrt{k}} + {\sqrt{k}}(i\partial_k)^l{\sqrt{k^3}}}
  \label{V^2}
\end{equation}
where $h^{(l)}(y) := [\partial^l_{y} f^2](y)$.

Both the formulas \eqref{expectationV} and \eqref{V^2} will be next to determine the quantum trajectories (both expectation values and variances) for the prescriptions described in this article, with a particular focus on the Yiang-Ding-Ma prescription.

\section{Quantum trajectories}
\label{sec:trajectories}

Given the form of a function $f$ (known once a particular prescription is selected), the results \eqref{expectationV} and \eqref{V^2} allow us to determine (up to corrections of the order $O(e^{-\braket{k}\pi})$ both the trajectory and variance of $V(\phi)$ as an explicit function of $\phi$ and two sequences of expectation values of operators, that are manifestly Dirac observables (thus correspond to constants of motion).  However, as the expectation values of these observables may grow with $l$, it is a priori not known whether under reasonable semiclassicality conditions the series on the righthand side of \eqref{expectationV}, \eqref{V^2} do converge. Therefore, it is beneficial to reexpress the expectation values of discussed Dirac observables via quantities more adapted to the notion of semiclassicality. A very useful example of such quantities are the central moments in the so-called Hamburger decomposition -- a central component of the \emph{semiclassical effective dynamics} \cite{Bojowald:2005cw}.

\subsection{Semiclassical form of constants of motion}
\label{sec:semiclass-eom}

Our point of departure here is an observation, that the operators acting in $k$-representation as $\hat{k}$ and $i\partial_k$ form (modulo $\hbar$ factor) a Heisenberg algebra, as $[\hat{k},i\partial_k] = -i\mathbb{I}$. Therefore, it is possible to define for them a set of central moments introduced in Appendix~\ref{sec:central}. It is straightforward to check by inspection that the rescaling of the structure constant in the algebra does affect neither the construction \eqref{gab1}, \eqref{eq:centr-inv} of the moments, nor the decomposition of the composite operators \eqref{eq:O-dec}, though it will affect the Poisson brackets between the moments. Given that, let us define the \emph{auxiliary moments} $\tilde{G}^{ab}$ as follows
\begin{equation}\label{eq:tGab-def}
  \tilde{G}^{a,b} := \sum_{k=0}^a\sum_{l=0}^b (-1)^{(a+b)-(k+l)}\binom{a}{k}\binom{b}{l} \braket{\hat{k}}^{a-k} \braket{i\partial_k}^{b-l} \tilde{F}^{k,l} \ , 
  \qquad
  \tilde{F}^{a,b}:=\braket{\hat{k}^a (i\partial_k)^b} \ .
\end{equation}
We then will attempt to express the expectation values of composite operators present in the expressions for $\braket{\tilde{V}^+_{\phi}}$ \eqref{generalformofvolumoperator} and $\braket{(\tilde{V}^+_{\phi})^2}$ \eqref{V^2}. Because if their slightly different form this step will be performed for the volume and its square separately. However, at this step we have to strengthen our restrictions on the studied semiclassical states, further requiring that:
\begin{enumerate}
  \item All the expectation values $\braket{\hat{k}}$, $\braket{i\partial_k}$, $\tilde{F}^{a,b}$ are finite. \item The state is peaked at large ``energy'' $\braket{\hat{k}}\gg 1$.
  \item On top of the above, the expectation values of the negative powers of $k$ must also remain finite:
  $\forall_{n\in\mathbb{N}}\ \braket{\widehat{k^{-n}}}<\infty$. 
\end{enumerate}
The reason for the last condition will become apparent further in this section. Note, that all the expectation values involved in these conditions are constants of motion, thus the properties required will not be lost in the process of dynamical evolution.

\subsubsection{Volume}

Let us focus our attention on the volume trajectory described via formula \eqref{expectationV}. In order to express its righthand side by central moments, we first need to express the operators $\sqrt{k}(i\partial_k)^l\sqrt{k}$ in terms of the Weyl-ordered ones.
In the first step, we commute the operators $\sqrt{k}$ and $(i\partial_k)^l$ in a symmetric fashion in order to bring the square roots together into integer powers
\begin{equation}\label{eq:V-conv}\begin{split}
  {\sqrt{k}}(i\partial_k)^l{\sqrt{k}}
  &= \frac{1}{2}\left(k(i\partial_k)^l+\sqrt{k}[(i\partial_k)^l,\sqrt{k}]+(i\partial_k)^lk-[(i\partial_k)^l,\sqrt{k}]\sqrt{k}\right)  \\
  &=\frac{1}{2}\Big(k(i\partial_k)^l+(i\partial_k)^lk\Big)+\frac{1}{2}[\sqrt{k},[(i\partial_k)^l,\sqrt{k}]]
  \ .
\end{split}\end{equation}
by comparing the 1st term in the 2nd line with McCoy formula for Weyl-ordered (further denoted by $\Weyl{\cdot}$) product operator
\begin{equation}
  X^nP^k=\frac{1}{2^n}\sum_{i=0}^n\binom{n}{i}X^iP^kX^{n-i}
\end{equation}
for $X=\hat{k}$, $Y=i\partial_k$ we get
\begin{equation}\label{sqrtkidpoksqrtk}
  {\sqrt{k}}(i\partial_k)^l{\sqrt{k}}
  =\, \Weyl{k(i\partial_k)^l} + \frac{1}{2}[\sqrt{k},[(i\partial_k)^l,\sqrt{k}]] \ .
\end{equation}

In the next step, we need to deal with the double commutator appearing above. In order to do that, we recall the (proven in \cite{2005JMP....46f3510T}) useful formula allowing us to expand the commutator involving function of one of the terms
\begin{equation}\label{eq:f-expansion}
  [P^l,f(X)]=\sum_{k=1}^l(-1)^{k+1}\binom{l}{k}[P,X]^k P^{l-k}f^{(k)}(X) \ .
\end{equation}
By applying it to both $\sqrt{k}$ terms in \eqref{sqrtkidpoksqrtk} we get
\begin{equation}
  [\sqrt{k},[(i\partial_k)^l,\sqrt{k}]]=\sum_{a=1}^l\sum_{b=1}^{l-a}(-1)^{(a+b)}\binom{l}{a}\binom{l-a}{b}i^{a+b}(i\partial_k)^{l-a-b}(\partial_k^a\sqrt{k})(\partial_k^b\sqrt{k})
    \label{doublecommutator}
\end{equation}
At this moment we note, that all the terms in the above expansion will have operators being products of natural powers of $i\partial_k$ and negative integer powers of $\hat{k}$. Since we are considering semiclassical states peaked at large $k$ we can estimate the expression by its leading term
\begin{equation}
  [\sqrt{k},[(i\partial_k)^l,\sqrt{k}]]
  = \frac{l(l-1)}{8}\left((i\partial_k)^{l-2}\frac{1}{k}+\frac{1}{k}(i\partial_k)^{l-2}\right)+O(\braket{k}^{-2}) \ .
  \label{doublecommutatorinVwitho(k)}
\end{equation} 
Treating $\widehat{1/k}$ as a composite operator one could convert it to a series involving $\braket{k}$ and moments $G^{0n}$, however, the relation \eqref{eq:O-dec} cannot be immediately applied to the whole leading term of \eqref{doublecommutatorinVwitho(k)} as it is not Weyl-ordered. Thus, we resort to a slightly more ``primitive'' approach: rewrite $\widehat{1/k}$ as 
\begin{equation}\label{eq:kneg-exp}
  \widehat{\frac{1}{k}} = \widehat{\frac{1}{k_0+(k-k_0)}} 
  = \sum_{n=0}^{\infty} \frac{(-1)^n}{k_0^{n+1}} (\hat{k} -k_0\mathbb{I})^n \ ,
\end{equation}
for some chosen constant $k_0$. Provided, that the expectation value $\braket{k}$ of the state is sufficiently close to $k_0$ the above series will be convergent. Plugging this back into \eqref{doublecommutatorinVwitho(k)} and grouping the terms by the power of $k$ we get
\begin{equation}
    \frac{1}{2}\braket{[\sqrt{k},[(i\partial_k)^l,\sqrt{k}]]}=\frac{l(l-1)}{16}\frac{1}{k_0}\sum_{n=0}^{\infty}\sum_{a=0}^n\frac{(-1)^{n+a}}{k_0^{n-a}}\binom{n}{a}\braket{(i\partial_k)^{l-2}k^{n-a}+k^{n-a}(i\partial_k)^{l-2}}
    \label{doublecommutatorwitho-k}
\end{equation}
which subsequently can be converted to Weyl ordering by the rule \eqref{simple}, giving
\begin{subequations}\begin{align}
  \frac{1}{2}\braket{[\sqrt{k},[(i\partial_k)^l,\sqrt{k}]]}
  &= \frac{l(l-1)}{8}\frac{1}{k_0}\sum_{n=0}^{\infty}\sum_{a=0}^n\sum_{j=0}^{\min(n-a,l-2)}\frac{1}{k_0^{n-a}}z(n,a,l,j)\braket{k^{n-a-2j}(i\partial_k)^{l-2-2j}}_{\rm Weyl} \ , 
  \\
  z(n,a,l,j)
  &:= (-1)^{n+a}\binom{n}{a}\frac{(i/2)^{2j}}{(2j)!}\frac{(n-a)!}{(n-a-2j)!}\frac{(l-2)!}{(l-2-2j)!} \ .
\end{align}\end{subequations}
At this moment we can apply the rule \eqref{eq:O-dec} (for expressing the composite operators in terms of the central moments) to the expectation values under the sum. Applying the result of this step to \eqref{eq:V-conv}, then \eqref{expectationV} we finally get
\begin{equation}\label{eq:Vexp-temp1}\begin{split}
  \braket{\hat{V_{\phi}}}
  &= A\sum_{l=0}^{\infty}\sum_{i=0}^l\frac{1}{i!}\frac{1}{(l-i)!}f^{(l)}(\beta\phi)\braket{i\partial_k}^{l-i}\Big(\braket{k}\Tilde{G}^{0,i}+\Tilde{G}^{1,i}\Big)
  \\
  &+ \frac{A}{8k_0}\sum_{l=2}^{\infty}\sum_{n=0}^{\infty}\sum_{a=0}^n\sum_{j=0}^{\min(n-a,l-2)}\sum_{b=0}^{n-a-2j}\sum_{c=0}^{l-2-2j} B_{abcjln}(\braket{k},\braket{i\partial_k},k_0) \Tilde{G}^{b,c} \ ,
\end{split}\end{equation}
where the 1st set of terms comes from Weyl-ordered term in \eqref{eq:V-conv} and the coefficients
$B_{abcjln}(\braket{k},\braket{i\partial_k},k_0)$ are
\begin{equation}
  B_{abcjln}(\braket{k},\braket{i\partial_k},k_0) := \frac{(i/2)^{2j}}{(2j)!}\frac{n!}{a!b!c!}\frac{(-1)^{n+a}\braket{k}^{n-a-2j-b}}{k_0^{n-a}(n-a-2j-b)!}\frac{f^{(l)}(\beta\phi)\braket{i\partial_k}^{l-2-2j-c}}{(l-2-2j-c)!}
\end{equation} 

Now we would like to change the order of summation in \eqref{eq:Vexp-temp1} in order to group terms involving the same $G$ moments. For the first terms, it is straightforward. In the second one, the desired order is: $bcjlna$. First we note that summation over $l$ commutes with summation over $n$ and $a$, the same happens between $b$ and $c$. Furthermore, by inspection, within the investigated sum we have $\min(n-a,l-2) = n-a$. Given that, by performing a set of consecutive translations (exchange of neighbouring summations), we can rewrite the sum as
\begin{equation}\label{eq:V-sum-reorder}\begin{split}
  &\sum_{l=2}^{\infty}\sum_{n=0}^{\infty}\sum_{a=0}^n\sum_{j=0}^{\min(n-a,l-2)}\sum_{b=0}^{n-a-2j}\sum_{c=0}^{l-2-2j} B_{abcjln}(\braket{k}, \braket{i\partial_k},k_0) \Tilde{G}^{b,c}  \\
  &= \sum_{b=0}^{\infty}\sum_{c=0}^{\infty}\sum_{j=0}^{\infty}\sum_{l=2+2j+c}^{\infty}\sum_{n=2j+b}^{\infty}\sum_{a=0}^{n-2j-b} B_{abcjln}(\braket{k}, \braket{i\partial_k},k_0) \Tilde{G}^{b,c} \ .
\end{split}\end{equation}
Let us now sum over $n$ and $a$ those terms in $B$ that depend on them. By (consecutively) substituting $m=n-2j-b$, regrouping the terms so that we distinguish a binomial $\binom{m}{a}$ and the terms not depending on particular summation index are brought outside the sum, applying the binomial theorem and setting $k_0\to\braket{k}$ we get
\begin{equation}\label{eq:Vsum-simpl}\begin{split}
    &\sum_{n=2j+b}^{\infty}\sum_{a=0}^{n-2j-b}\frac{n!}{a!}\frac{(-1)^{n+a}\braket{k}^{n-a-2j-b}}{k_0^{n-a}(n-a-2j-b)!}
    = \frac{(-1)^{(2j+b)}}{k_0^{(2j+b)}}\sum_{m=0}^{\infty}\sum_{a=0}^{m}(-1)^{m+a}\frac{(m+2j+b)!}{a!(m-a)!}\frac{\braket{k}^{m-a}}{k_0^{m-a}}
    \\
    &= \frac{(-1)^{(2j+b)}}{k_0^{(2j+b)}}\sum_{m=0}^{\infty}\frac{(m+2j+b)!}{m!}\sum_{a=0}^{m}\binom{m}{a}(-1)^{m-a}\frac{\braket{k}^{m-a}}{k_0^{m-a}}=\frac{(-1)^{(2j+b)}}{k_0^{(2j+b)}}\sum_{m=0}^{\infty}\frac{(m+2j+b)!}{m!}(1-\frac{\braket{k}}{k_0})^m
    \\
    &= \frac{(-1)^{(2j+b)}}{\braket{k}^{(2j+b)}}(2j+b)! \ ,
\end{split}\end{equation}
Note that, while setting $k_0=\braket{k}$ is in general not possible in the semiclassical effective dynamics, it can be performed here as our basic operators are constants of motion and in consequence, we never need to probe the dynamics of central moments. By continuity of all the terms $(1-\braket{k}/k_0)^m$ as $k_0\to\braket{k}$ only those for $m=0$ contribute, giving the last equality above. 

In the next step, we perform the summation over $l$ those terms in $B$ that depend on it. Remarkably, they gather to Taylor series that correspond to certain derivatives of $f$ at the argument shifted by $\braket{i\partial_k}$, that is
\begin{align}
    \sum_{l=2+2j+c}^{\infty}\frac{f^{(l)}(\beta\phi)\braket{i\partial_k}^{l-2-2j-c}}{(l-2-2j-c)!}=f^{(c+2j+2)}(\beta\phi-\braket{i\partial_k})
\end{align}

By plugging both the results above back to \eqref{eq:Vexp-temp1} (with sums already reordered via \eqref{eq:V-sum-reorder}) we arrive to a final form of the (modified) volume trajectory
\begin{equation}\label{formulaforv}\begin{split}
  \braket{\tilde{V}^+_{\phi}} 
  &= A\sum_{l=0}^{\infty}\frac{1}{l!}f^{(l)}(\beta\phi-\braket{l\partial_k})\Big(\braket{k}\Tilde{G}^{0,l}+\Tilde{G}^{1,l}\Big)
  \\
  &+ \frac{A}{8\braket{k}}\sum_{b=0}^{\infty}\frac{(-1)^b}{b!}\sum_{c=0}^{\infty}\frac{1}{c!} \Tilde{G}^{b,c} \sum_{j=0}^{\infty}\frac{(-1)^j}{2^{2j}(2j)!}(2j+b)!f^{(c+2j+2)}(\beta\phi-\braket{i\partial_k})\braket{k}^{-2j-b} \ ,
\end{split}\end{equation}
Note, that all the terms in the 2nd line are proportional to the negative powers of $\braket{k}$, thus under our assumptions regarding considered class of states they are small corrections to be neglected in further analysis. Given that the corrections between $\braket{\hat{V}_{\phi}}$ and $\braket{\tilde{V}^+_\phi}$ are much smaller, we can write the final form of the (approximated) volume trajectory as
\begin{equation}\label{eq:Vexp-final}
  \braket{\hat{V}_\phi} = A\sum_{l=0}^{\infty}\frac{1}{l!}f^{(l)}(\beta\phi-\braket{l\partial_k})\Big(\braket{k}\Tilde{G}^{0,l}+\Tilde{G}^{1,l}\Big) + O(\braket{k}^{-1}) \ .
\end{equation}

\subsubsection{Volume squared and variance}

Let us now turn our attention to the trajectory of the squared volume operator $\braket{\hat{V}_{\phi}}$, a component needed to control the variance of the semiclassical state during the evolution. As in the case of the volume itself, we will use the approximate operator $\tilde{V}^+_{\phi}$. Our point of departure is formula \eqref{V^2}. The procedure of its conversion to the series involving the central moments is a repetition of the one already employed for the volume itself. Its first step is the transformation (analogous to the one leading to \eqref{sqrtkidpoksqrtk}) of the expectation values of $\braket{\sqrt{k^3}(i\partial_k)^l\sqrt{k}+\sqrt{k}(i\partial_k)^l\sqrt{k^3}}$ to those of Weyl ordered operator
\begin{equation}
  \frac{1}{2}\Big(\braket{{\sqrt{k^3}}(i\partial_k)^l{\sqrt{k}}+{\sqrt{k}}(i\partial_k)^l{\sqrt{k^3}}}\Big)
  = \braket{k^2(i\partial_k)^l}_{Weyl}+\frac{1}{2}\braket{[\sqrt{k},[(i\partial_k)^l,\sqrt{k}]]k+k[\sqrt{k},[(i\partial_k)^l,\sqrt{k}]]} \ ,
\end{equation}
and a double commutator remnant, which we expect to be a small correction. By $(i)$ applying the expansion \eqref{eq:f-expansion}, $(ii)$ expanding the negative powers of $\hat{k}$ around some chosen $k_0$ \eqref{eq:kneg-exp}, and $(iii)$ grouping the terms by monomials in $\hat{k}$, $i\partial_k$ we arrive to an analog of \eqref{doublecommutatorwitho-k}
\begin{equation}
  \frac{1}{2}\Big(\braket{{\sqrt{k^3}}(i\partial_k)^l{\sqrt{k}}}+\braket{{\sqrt{k}}(i\partial_k)^l{\sqrt{k^3}}}\Big)
  = \braket{k^2(i\partial_k)^l}_{Weyl}+\frac{l(l-1)}{8}\braket{(i\partial_k)^{l-2}}
  + O(\braket{k}^{-1}) \ .
  \label{scalarproductshowniv2}
\end{equation}
Plugging this result back into \eqref{V^2} and implementing: $(i)$ the formula \eqref{gab2}, $(ii)$ the process of summation reordering analogous to \eqref{eq:V-sum-reorder} and $(iii)$ the simplification upon setting $k_0\to\braket{k}$ we get an approximation
\begin{equation}\label{eq:V2-approx-final}\begin{split}
  \braket{(\tilde{V}^+_{\phi})^2}
  &= A^2\sum_{i=0}^{\infty}\sum_{a=0}^i\frac{1}{(i-a)!a!}f^{(i-a)}(\beta\phi-\braket{i\partial_k})f^{(a)}(\beta\phi-\braket{i\partial_k})(\braket{k}^2\Tilde{G}^{0,i}+2\braket{k}\Tilde{G}^{1,i}+\Tilde{G}^{2,i}) 
  \\
  &+ \frac{A^2}{8}\sum_{i=0}^{\infty}\sum_{a=0}^{i+2}\frac{(i+2)(i+1)}{(i+2-a)!a!}f^{(i+2-a)}(\beta\phi-\braket{i\partial_k})f^{(a)}(\beta\phi-\braket{i\partial_k})\Tilde{G}^{0,i} 
  + O(\braket{k}^{-1})\ ,
\end{split}\end{equation}
where the expectation value $\braket{(\tilde{V}^+_{\phi})^2}$ can be again replaced with $\braket{\hat{V}^2_{\phi}}$ as the difference between the two is of subleading order with respect to the correction on the righthand side of the above equation.

The variance of the state at given $\phi$ is given by the standard deviation formula
\begin{equation}\label{eq:variance-def}
    \sigma(V_{\phi})=\sqrt{\braket{V_{\phi}^2}-\braket{V_{\phi}}^2}
\end{equation}
to which we can directly plug \eqref{eq:V2-approx-final} and the square of \eqref{formulaforv} which in turn is of the form
\begin{equation}\begin{split}
  \braket{\tilde{V}^+_{\phi}}^2
  &= A^2\sum_{i=0}^{\infty}\sum_{i'=0}^{\infty}\frac{1}{i!}{\rm f}^{(i)}\frac{1}{i'!}{\rm f}^{(i')}\Big(\braket{k}^2\Tilde{G}^{0,i}\Tilde{G}^{0,i'}+2\braket{k}\Tilde{G}^{0,i}\Tilde{G}^{1,i'}+\Tilde{G}^{1,i}\Tilde{G}^{1,i'}\Big) 
  \\
  &+ \frac{A^2}{4}\sum_{i=0}^{\infty}\frac{1}{i!}\sum_{b=0}^{\infty}\frac{(-1)^b}{b!}\sum_{c=0}^{\infty}\frac{1}{c!}\sum_{j=0}^{\infty}\frac{(-1)^j}{2^{2j}(2j)!}(2j+b)!{\rm f}^{(i)}{\rm f}^{(c+2j+2)}\braket{k}^{-2j-b}\Tilde{G}^{b,c}\Tilde{G}^{0,i}
  \\
  &+ \frac{A^2}{4\braket{k}}\sum_{i=0}^{\infty}\frac{1}{i!}\sum_{b=0}^{\infty}\frac{(-1)^b}{b!}\sum_{c=0}^{\infty}\frac{1}{c!}\sum_{j=0}^{\infty}\frac{(-1)^j}{2^{2j}(2j)!}(2j+b)!{\rm f}^{(i)}{\rm f}^{(c+2j+2)}\braket{k}^{-2j-b}\Tilde{G}^{b,c}\Tilde{G}^{1,i}
  \\
  &+ \frac{A^2}{64\braket{k}^2}\sum_{b,b'}^{\infty}\sum_{c,c'}^{\infty}\sum_{j,j'}^{\infty}\frac{(-1)^{j+j'+b+b'}(2j+b)!(2j'+b')!}{2^{2(j+j')}b!b'!c!c'!(2j)!(2j')!}\frac{{\rm f}^{(c+2j+2)}{\rm f}^{(c'+2j'+2)}}{\braket{k}^{(2j+b+2j'+b')}}\Tilde{G}^{b,c}\Tilde{G}^{b',c'} 
\end{split}\end{equation}
(where ${\rm f}^{(a)} := f^{(a)}(\beta\phi-\braket{i\partial_k})$) and of which the last two lines are a remnant of the order $O(\braket{k}^{-1}$. As a result, the squared variance $\sigma^2(\hat{V}_{\phi}) = \braket{\hat{V}^2_{\phi}} - \braket{\tilde{V}^+_{\phi}}^2 + O(e^{-\braket{k}\pi})$ takes the form
\begin{equation}\begin{split}
  \sigma^2(V_{\phi})
  = A^2\sum_{i=0}^{\infty}\Big(\braket{k}^2&\big(
    \sum_{a=0}^i\frac{1}{(i-a)!a!}{\rm f}^{(i-a)}{\rm f}^{(a)}\Tilde{G}^{0,i}-\sum_{i'=0}^{\infty}\frac{1}{i!}\frac{1}{i'!}{\rm f}^{(i)}{\rm f}^{(i')}\Tilde{G}^{0,i}\Tilde{G}^{0,i'}\big) 
    \\
    +2\braket{k} &\big(
    \sum_{a=0}^i\frac{1}{(i-a)!a!}{\rm f}^{(i-a)}{\rm f}^{(a)}\Tilde{G}^{1,i}-\sum_{i'=0}^{\infty}\frac{1}{i!}\frac{1}{i'!}{\rm f}^{(i)}{\rm f}^{(i')}\Tilde{G}^{0,i}\Tilde{G}^{1,i'}\big)
    \\
    + &\big(
    \sum_{a=0}^i\frac{1}{(i-a)!a!}{\rm f}^{(i-a)}{\rm f}^{(a)}\Tilde{G}^{2,i}-\sum_{i'=0}^{\infty}\frac{1}{i!}\frac{1}{i'!}{\rm f}^{(i)}{\rm f}^{(i')}\Tilde{G}^{1,i}\Tilde{G}^{1,i'}\big)
    \\
    +&\big(
    \frac{1}{8}\sum_{a=0}^{i+2}\frac{(i+2)(i+1)}{(i+2-a)!a!}{\rm f}^{(i+2-a)}{\rm f}^{(a)}\Tilde{G}^{0,i}-\frac{1}{4}\sum_{i'=0}^{\infty}\frac{1}{i!}\frac{1}{i'!}{\rm f}^{(i)}{\rm f}^{(i'+2)}\Tilde{G}^{0,i}\Tilde{G}^{0,i'}\big)\Big) \\
    + O(\braket{k}^{-1}&)  \ .
  \label{standarddeviation}
\end{split}\end{equation}
Note, that the terms in the last two lines will give a contribution to the variance itself \eqref{eq:variance-def}, that will be of the order $O(\braket{k}^{-1}$, thus under the approximations taken they can again be neglected.

At this point, we have at our disposal the trajectory of volume at given $\phi$ (of a general form given by \eqref{eq:V-gen}) and its variance given up to corrections $O(\braket{k}^{-1})$ expressed in terms of the explicit functions of $\phi$, expectation values $\braket{k}$, $\braket{i\partial_k}$ and the auxiliary central moments $\tilde{G}^{a,b}$. We will now apply them to particular prescriptions discussed in this article.

\subsection{Applications}
\label{sec:V-applications}

Before proceeding with applying the results of the previous subsection we need to reexamine briefly our construction of central moments. Since the commutator between $\hat{k}$ and $i\partial_k$ is not proportional to $\hbar$, the moments $\tilde{G}^{ab}$ do not reproduce the usual hierarchy of corrections, where each order scales with appropriate power of $\hbar$. Also, with $k$ being dimensionless, the physical interpretation of the operators is not immediately obvious. These issues are however easy to fix. First, we note that the action of the ``energy'' (the scalar field momentum) operator $\hat{p}_{\phi} = \sqrt{\Theta}$ \eqref{Diracsmlqc2} in $k$-representation takes a simple form $\hat{p}_{\phi} = \hbar\beta\hat{k}$. On the other hand, the operator $i\beta^{-1}\partial_k$ is the canonical momentum of $\hat{p}_{\phi}$ and the form in which $\braket{i\partial_k}$ enters the expressions describing volume trajectory and variance indicates, that $i\beta^{-1}\partial_k$ can be interpreted as certain shift (offset) in $\phi$. Let us then denote it by $\hat{\phi}_0$. Consequently, our canonical pair of basic observables will be
\begin{equation}\label{eq:phys-var}
  \hat{p}_{\phi} = \hbar\beta\hat{k} \ , \qquad
  \hat{\phi}_0 := i\beta^{-1}\partial_k \ , \qquad 
  [\hat{\phi}_0,\hat{p}_{\phi}] = i\hbar\mathbb{I} \ .
\end{equation}
For this pair, we can now build a set of central moments via \eqref{gab1}. Comparing it with the definition of the auxiliary ones \eqref{eq:tGab-def} we see, that they are related in a simple way
\begin{equation}\label{eq:phys-G}
  G^{ab} = \hbar^a \beta^{a-b} \tilde{G}^{a,b} \ ,
\end{equation}
which allows to easily convert the results of this section. In particular, the trajectory of the volume \eqref{eq:Vexp-final} in new variables reads
\begin{equation}
  \braket{\hat{V_{\phi}}} 
  = \frac{A}{\hbar}\sum_{i=0}^{\infty}\frac{\beta^{i-1}}{i!}f^{(i)}(\beta(\phi-\phi_0))\Big(p_\phi G^{0,i}+G^{1,i}\Big) + O(p_{\phi}^{-1}) \ , 
\label{finalformforexpV}
\end{equation}
where $p_{\phi} := \braket{\hat{p}_{\phi}}$ and $\phi_0 := \braket{\hat{\phi}_0}$. Similarly, the square of the variance \eqref{standarddeviation} takes the form
\begin{equation}\label{eq:sigma2-final}\begin{split}
  \sigma^2(V_{\phi})
  = \frac{A^2}{\hbar^2}\sum_{i=0}^{\infty}\Big(p_{\phi}^2&\big(
    \sum_{a=0}^i\frac{\beta^{i-2}}{(i-a)!a!}{\rm f}^{(i-a)}{\rm f}^{(a)}G^{0,i}-\sum_{i'=0}^{\infty}\frac{\beta^{i+i'-2}}{i!i'!}{\rm f}^{(i)}{\rm f}^{(i')}G^{0,i}G^{0,i'}\big) 
    \\
    +2p_{\phi} &\big(
    \sum_{a=0}^i\frac{\beta^{i-2}}{(i-a)!a!}{\rm f}^{(i-a)}{\rm f}^{(a)}G^{1,i}-\sum_{i'=0}^{\infty}\frac{\beta^{i+i'-2}}{i!i'!}{\rm f}^{(i)}{\rm f}^{(i')}G^{0,i}G^{1,i'}\big)
    \\
    + &\big(
    \sum_{a=0}^i\frac{\beta^{i-2}}{(i-a)!a!}{\rm f}^{(i-a)}{\rm f}^{(a)}G^{2,i}-\sum_{i'=0}^{\infty}\frac{\beta^{i+i'-2}}{i!i'!}{\rm f}^{(i)}{\rm f}^{(i')}G^{1,i}G^{1,i'}\big)
    \\
    +&\hbar^2\big(
    \frac{1}{8}\sum_{a=0}^{i+2}\frac{(i+2)(i+1)\beta^{i-2}}{(i+2-a)!a!}{\rm f}^{(i+2-a)}{\rm f}^{(a)}G^{0,i}-\frac{1}{4}\sum_{i'=0}^{\infty}\frac{\beta^{i+i'-2}}{i!i'!}{\rm f}^{(i)}{\rm f}^{(i'+2)}G^{0,i}G^{0,i'}\big)\Big) \\
    + O(p_{\phi}^{-1}&)  \ .
\end{split}\end{equation}
Note, that the terms in the $4$th line will give a nontrivial contribution also when we set all the moments $G^{a,b}$ for $a+b\geq 2$ to zero, which corresponds to partial classical limit. In that case we have
\begin{equation}
  \sigma^2_0(V_{\phi}) = \frac{A^2}{4\beta^2} \left( f^{(1)}(\beta(\phi-\phi_0)) \right)^2 
  + O(p_{\phi}^{-1}) \ ,
\end{equation}
however upon applying the same partial limit to \eqref{finalformforexpV} (of which result we denote as $V_0(\phi)$) we get
\begin{equation}\label{eq:sigma-free}
  \frac{\sigma^2_0(V_{\phi})}{V_0^2(\phi)} 
  = \frac{\hbar^2}{4} \left( \frac{f'(\beta(\phi-\phi_0))}{f(\beta(\phi-\phi_0))} \right)^2 
  + O(p_{\phi}^{-1}) \ ,
\end{equation}
thus $\sigma_0(V_{|phi})$ is suppressed by $\hbar$ with respect to the volume. Furthermore, in the case of the actual quantum evolution of states satisfying the assumptions stated at the beginning of this section (where variances do not vanish), $\sigma^2(V_{\phi})$ will be dominated by (nonvanishing) terms proportional to $p_{\phi}^2$, thus the contribution of the ``free'' (not multiplied by any central moment of 2nd and higher order) to the variance $\sigma(V_{\phi})$ itself will be of the order $O(p_{\phi}^{-1})$ which we neglected in our approximation.

These results are valid for a wide class of models -- in general, they can be directly applied to those, for which $(i)$ the Hamiltonian constraint can be brought to a Klein-Gordon form \eqref{eq:Ctot_q}, where $(ii)$ the evolution operator can again be brought by introducing an appropriate coordinate to a form of second order derivative and $(iii)$ for which the volume operator can be expressed in the form \eqref{eq:V-gen}. As mentioned earlier, in the context of a flat FRLW Universe with a massless scalar field this applies in particular to the mainstream LQC and Yang-Ding-Ma prescription, as well as to a geometrodynamical analog of the model. One has to remember, however, that the derivation of \eqref{finalformforexpV} and \eqref{eq:sigma2-final} employs a semiclassical treatment, which in turn is sensitive to the existence of sufficiently large sector semiclassical in selected variables. This is not always the case. One such counterexample is the prescription based on strict Thiemann regularization, which we will also briefly discuss below.
Let us start with the least investigated Yang-Ding-Ma prescription.

\subsubsection{Yang-Ding-Ma prescription}

In this prescription, the action of the volume operator is given in \eqref{Vinmlqc2inx}, which is a specific case of \eqref{eq:V-gen} with
\begin{equation}\label{eq:Af-YDM}
  A = \frac{\alpha}{\sqrt{\gamma^2+1}} \ , \qquad
  f(x) = (\gamma^2\tanh^2(x)+1)\cosh(x) = (\gamma^2+1)\cosh(x)-\frac{\gamma^2}{\cosh(x)} \ .
\end{equation}
Plugging these forms of $A$, $f$ into \eqref{finalformforexpV} we arrive to an expression for the volume trajectory
\begin{equation}\label{finalformforexpVmlqc2}\begin{split}
  \braket{\hat{V_{\phi}}}
  &= \frac{\gamma\sqrt{\pi G\Delta}}{\sqrt{3(\gamma^2+1})} p_{\phi} \left(\gamma^2\tanh^2(\beta(\phi-\phi_0))+1\right)\cosh(\beta(\phi-\phi_0))
  \\
  &+ \frac{2\pi\gamma G\sqrt{\Delta}}{\sqrt{\gamma^2+1}} \left( (\gamma^2+1)\sinh(\beta(\phi-\phi_0)) + \gamma^2\frac{\sinh(\beta(\phi-\phi_0))}{\cosh^2(\beta(\phi-\phi_0))} \right) G^{1,1}
  \\
  &+ \frac{\gamma(\pi G)^{3/2}\sqrt{\Delta}}{\sqrt{\gamma^2+1}} \left( (\gamma^2+1)\cosh(\beta(\phi-\phi_0)) - \gamma^2\frac{\cos^2(\beta(\phi-\phi_0))-2}{\cos^3(\beta(\phi-\phi_0))} \right) \left( p_{\phi}G^{0,2} + G^{1,2} \right)
  \\
  &+ \frac{2\pi\gamma G\sqrt{\Delta}}{\sqrt{\gamma^2+1}} \sum_{i=3}^{\infty}\frac{(12\pi G)^{(i-1)/2}}{i!} f^{(i)}(\beta(\phi-\phi_0))\left(p_gG^{0,i}+G^{1,i}\right)
  + O(p_{\phi}^{-1}) \ ,
\end{split}\end{equation}
where we distinguished the corrections of up to 2nd order, furthermore explicitly evaluating the first two derivatives of $f$ from the rightmost expression in \eqref{eq:Af-YDM}.

Remarkably, the term in the 1st line of the above equation (corresponding to 0th order semiclassical approximation) agrees with the trajectory derived heuristically in \cite{Li:2018fco}. The 2nd order corrections are proportional to the variance of $\hat{\phi}_0$ --$G^{0,2}$-- and the correlation $G^{1,1}$ out of which the 1st one will be dominant as being further multiplied by $p_{\phi}$. The variance of $p_{\phi}$ does not influence the trajectory, which is a general feature visible already in \eqref{finalformforexpV}.

Similarly, we can evaluate the variance, plugging $A$ and $f$ into \eqref{eq:sigma2-final}. This yields
\begin{equation}\label{stdmlqc2}\begin{split}
  \sigma^2(V_{\phi}) &= \frac{A^2}{\hbar^2} p_{\phi}^2 \left( (\gamma^2+1)\sinh(\beta(\phi-\phi_0)) + \gamma^2\frac{\sinh(\beta(\phi-\phi_0))}{\cosh^2(\beta(\phi-\phi_0))} \right)^2 G^{0,2} 
  \\
  &+ \frac{A^2}{\hbar^2} 2p_{\phi} \left( (\gamma^2+1)\sinh(\beta(\phi-\phi_0)) + \gamma^2\frac{\sinh(\beta(\phi-\phi_0))}{\cosh^2(\beta(\phi-\phi_0))} \right) \left( (\gamma^2+1)\cosh(x)-\frac{\gamma^2}{\cosh(x)} \right) G^{1,1} \\
  &+ \delta\sigma^2(V_{\phi}) + O(p_{\phi}^0) \ ,
\end{split}\end{equation}
where $\delta\sigma^2(V_{\phi})$ depends on corrections of the $3rd$ and higher order and $O(p_{\phi})$ will give to the variance $\sigma(V_{\phi})$ a correction of the order $O(p_{\phi}^{-1})$, neglected under approximations taken. We see that, just like in case of the corrections to $\braket{\hat{V}_{\phi}}$ the dominant contribution to $\sigma(V_{\phi}$ will come from the variance of $\hat{\phi}_0$. 

Let us now reexamine the ``free'' term \eqref{eq:sigma-free}. While it is suppressed by $\hbar$, it could in principle grow to macroscopic level if for some $\phi$ the ratio $f'(\beta(\phi-\phi_0))/f(\beta(\phi-\phi_0))$ becomes large. However, a direct inspection shows, that for $f$ given by \eqref{eq:Af-YDM} we have
\begin{equation}
  \forall x\in\re\ \left|\frac{f'(x)}{f(x)}\right| < 1.02 \ , \qquad
  \lim_{x\to\pm\infty} \frac{f'(x)}{f(x)} = \pm 1 \ .
\end{equation}
This excludes the possibility of an uncontrollable growth of the term under scrutiny.

Finally, in order to understand the role of the correlation $G^{1,1}$ let us compare the relative variances of volume at the distant future and past. At this time we will neglect the corrections of 3rd and higher order as well as the corrections of the order $O(p_{\phi}^{-1})$. Then, from \eqref{finalformforexpVmlqc2} and \eqref{stdmlqc2} and from the observation, that
\begin{equation}
  \lim_{x\to\pm\infty} f^{(n)}(x) e^{-|x|} = (\pm 1)^n \frac{\gamma^2+1}{2} \ ,
\end{equation}
it follows
\begin{equation}
  \left( \lim_{\phi\to\infty} - \lim_{\phi\to-\infty} \right) \frac{\sigma^2(V_{\phi})}{\braket{\hat{V}_{\phi}}^2} 
  \approx 2\beta^2 \frac{G^{1,1}}{p_{\phi}} \left( \lim_{\phi\to\infty} - \lim_{\phi\to-\infty} \right) \frac{f^{(1)}(\beta(\phi-\phi_0))}{f(\beta(\phi-\phi_0))} 
  = 4\beta^2 \frac{G^{1,1}}{p_{\phi}} \ . 
\end{equation}
As a consequence, the difference in squared relative variances of the volume in the distant future and past is (within the approximation implemented) proportional to the relative correlation $G^{1,1}/p_{\phi}$. 
Since due to the Schwartz inequality the correlation is bound by variances
\begin{equation}
  G^{2,0}G^{0,2} \geq \frac{1}{4} (G^{1,1})^2 \ ,
\end{equation}
the possible loss of semiclassicality between the distant future and past is severely restricted.

\subsubsection{Mainstream LQC}

For this prescription, originally formulated in \cite{Ashtekar:2006wn} the total Hamiltonian constraint is in the same form \eqref{eq:Ctot_q}, however the operator $\Theta$ in $v$-representation is a difference operator of $2$nd order and the sectors invariant under its actions are regular lattices in $v$ separated by $4$ instead of $2$. Barring this minor difference, the division onto the superselection sectors is the same and the usual studies focus on a single sector of symmetric functions supported on a lattice $v\in 4\mathbb{Z}$. Also here it is convenient to transform to $b$-representation (see \cite{Ashtekar:2007em} for the original implementation of this representation with slightly different variables), where $b$ is a canonical momentum defined through \eqref{eq:vb-def}, \eqref{eq:vb-Poisson}. The transformation between the representations is very similar to the one already discussed and will be again given by \eqref{eq:sym_vb} and \eqref{eq:vb-trans-inv} with one modification, that the summation in \eqref{eq:sym_vb} is performed over lattice $v\in 4\mathbb{Z}$ and the integration in the inverse transform \eqref{eq:vb-trans-inv} is performed over the range $b\in [0,\pi]$. The evolution operator is positive definite, essentially selfadjoint \cite{Kaminski:2007ew} and in $b$-representation the evolution operator $\Theta$ takes the form
\begin{equation}
    \Theta = -12\pi G\hbar^2 (\sin(b)\partial_b)^2 \ ,
\end{equation}
thus becomes $2$nd order differential operator. Introducing auxiliary function $x:=\ln(\tan(b/2))$ we bring $\Theta$ to an explicit Klein-Gordon form, namely
\begin{equation}
  \Theta = -12\pi G\hbar^2 \partial_x^2 \ .
\end{equation}

The physical Hilbert space is spanned by the energy eigenbasis
\begin{equation}
  e_k{x} = N_k\cos(kx) \ , [\Theta e_k](x) = 12\pi G\hbar e_k(x) = \beta^2\hbar^2 e_k(x) \ ,
\end{equation}
where by considerations analogous to the ones in Appendix~\ref{sec:WDW-limit} (see in particular \cite{Bodendorfer:2018zyr}) the normalization factor is
\begin{equation}
  N_k = \frac{1}{\sqrt{k}} \left( 1 + O(e^{-k\pi/2}) \right) \ .
\end{equation}
Similarly, the physical inner product can again be approximated (up to a remnant $O(e^{-\braket{k}\pi})$) by \eqref{scalarxapproximate} with the projections onto left/righthand moving plane wave components given again by \eqref{eq:proj-final}.

The observable corresponding to the volume in $x$-representation takes the form of \eqref{eq:V-gen} with 
\begin{equation}\label{eq:const-sLQC}
  A = \alpha = 2\pi\gamma G\hbar \sqrt{\Delta} \ , \qquad
  f(x) = \cosh(x) \ ,
\end{equation}
and can again be approximated by the operator $\tilde{V}^+$ given in \eqref{generalformofvolumoperator} (with the same correction order), thus the construction of semiclassical moments and the results of sec.~\ref{sec:semiclass-eom}, as well as the initial (non-prescription specific) part of sec.~\ref{sec:V-applications} can be applied here. In particular, the trajectory of the volume in $\phi$ (calculated by plugging \eqref{eq:const-sLQC} into \eqref{finalformforexpV}) can be explicitly calculated up to an arbitrary semiclassical quantum correction order
\begin{equation}\label{finalformforexpVlqc}\begin{split}
  \braket{\hat{V_{\phi}}}
  &= \frac{\gamma\sqrt{\pi G\Delta}}{\sqrt{3}}p_{\phi}\cosh(\beta(\phi-\phi_0))+\frac{\gamma\sqrt{\pi G\Delta}}{\sqrt{3}}\sinh(\beta(\phi-\phi_0))\sum_{i=1}^{\infty}\frac{\beta^{2i-1}}{(2i-1)!}\Big(p_{\phi}G^{0,2i-1}+G^{1,2i-1}\Big)
  \\
  &+ \frac{\gamma\sqrt{\pi G\Delta}}{\sqrt{3}}\cosh(\beta(\phi-\phi_0))\sum_{i=1}^{\infty}\frac{\beta^{2i}}{(2i)!}\Big(p_{\phi}G^{0,2i}+G^{1,2i}\Big)
  + O(p_{\phi}^{-1})  \\
  &= \frac{\gamma\sqrt{\pi G\Delta}}{\sqrt{3}} p_{\phi} C_V \cosh(\beta(\phi-(\phi_0+\delta\phi)))  
  + O(p_{\phi}^{-1}) \ ,
\end{split}\end{equation}
where the constants
\begin{subequations}\begin{align}
  C_V
  &= \sqrt{\left(1+\sum_{i=1}^{\infty}\frac{\beta^{2i}}{(2i)!}\Big(G^{0,2i}+\frac{G^{1,2i}}{p_{\phi}}\Big)\right)^2-\left(\sum_{i=1}^{\infty}\frac{\beta^{2i-1}}{(2i-1)!}\Big(G^{0,2i-1}+\frac{G^{1,2i-1}}{p_{\phi}}\Big)\right)^2} 
  \\
  \delta\phi 
  &= \frac{1}{\beta}\tanh^{-1}\left( \left( \sum_{i=1}^{\infty}\frac{\beta^{2i-1}}{(2i-1)!}\Big(G^{0,2i-1}+\frac{G^{1,2i-1}}{p_{\phi}}\Big) \right) \Big/ \left(1+\sum_{i=1}^{\infty}\frac{\beta^{2i}}{(2i)!}\Big(G^{0,2i}+\frac{G^{1,2i}}{p_{\phi}}\Big)\right) \right) \ ,
\end{align}\end{subequations}
correspond to relative change due to quantum corrections of the bounce volume and the shift of the bounce (scalar field) time respectively.

Similarly, (plugging \eqref{eq:const-sLQC} into \eqref{eq:sigma2-final}) one can express the square of the variance in volume up to arbitrary order, though the expression will be quite long. Thus, for compactness we will expand only the $2$nd order terms, just like in \eqref{stdmlqc2}, leaving the higher order corrections as $\delta\sigma^2(V_{\phi})$
\begin{equation}
  \sigma^2(V_{\phi})
  = 4\pi^2\gamma^2G^2\Delta \left( p_{\phi}^2 \sinh^2(\beta(\phi-\phi_0)) G^{0,2} + p_{\phi} \sinh(2\beta(\phi-\phi_0)) G^{1,1} \right) + \delta\sigma^2(V_{\phi}) + O(p_{\phi}^0) \ .
   \label{standard deviation mlqc2}
\end{equation}
The potentially problematic free term \eqref{eq:sigma-free} remains suppressed also in this prescription as $f'(x)/f(x) = \tanh(x) \in [-1,1]$.

\subsubsection{Wheeler-DeWitt analog}

The applicability of the trajectory evaluation method introduced in this article is not restricted to just the models based on polymer quantization and in particular can be easily applied to the geometrodynamical (also known as Wheeler-DeWitt) quantization of the cosmological model studied here. We briefly recall its main properties in Appendix.~\ref{sec:WDW}. Before proceeding, let us note however, that in this model the evolution operator can be easily written in the Klein-Gordon form without shifting to the representation momentum $b$ (with transformation now defined through \eqref{eq:vb-trans-WDW}). Indeed, via a simple rescaling of physical states by a fixed function of $v$ and introducing for symmetric states the variable $y=\ln|v|$ we bring $\ub{\Theta}$ again to the form $\ub{\Theta} = 12\pi G\hbar^2\partial_y^2$. In this variable the volume operator will read $V = \alpha\exp(y)$ thus the problem of finding the trajectories reduces to a textbook one. Despite that, we will proceed with working in $b$ representation as: $(i)$ it will serve as a test for out method in a new setting, and $(ii)$ it may be more convenient to apply than the volume one in some more complicated models (like for example the ones admitting nonvanishing cosmological constant, see \cite{Bentivegna:2008bg,Pawlowski:2011zf}). 

The transformation to $b$-representation, the form of the evolution operator and the energy eigenbasis are presented in detail in Appendix.~\ref{app:WDW-b}. The trajectory extraction procedure is a repetition of the one used in the previous sub-subsection, however there are two crucial differences affecting its detailed implementation:
\begin{enumerate}
  \item If the requirement that the state is symmetric in $v$ (thus in $b$) is dropped, the transformation    bringing the evolution operator 
    \begin{equation}
      \ub{\Theta} = -12\pi G \hbar^2 (b\partial_b)^2 \ ,
    \end{equation}
    to the explicit Klain-Gordon form needs to use two copies of real coordinate:
    \begin{equation}
      \re\ni x^{\pm} = \ln(\pm b) \ .
    \end{equation}
    Fortunately, upon restricting to the symmetric states only one coordinate $x:=\ln|b|$ suffices.
  \item Upon restriction to symmetric states, the energy eigenbasis is twofold degenerate (see
    \eqref{eq:WDW-state}-\eqref{eq:WDW-disp}) and the variable (analog of a wave number) $k$ spans the whole real line. The sign of $k$ now distinguishes lefthand moving (or contracting in $v$) for $k<0$ and righthand moving (or expanding in $v$) plane waves. In $b$-representation they correspond to (respectively) righthand/lefthand moving plane waves \eqref{eq:WDW-basis-b}, not the standing waves like in LQC models considered earlier. In consequence, the approximation of the volume by \eqref{generalformofvolumoperator}, crucial for simplifying the calculations, will not hold in general.
\end{enumerate}
The first property implies, that working with variable $x$ requires to pick one sign of $b$. 
In order to deal with the second problem we note, that the (specified at the beginning of sec.~\ref{sec:semiclass-eom}) assumptions regarding the semiclassical states we are working with imply in particular, that the bulk of the considered state is necessarily located on \emph{one} orientation (sign) of $k$, with the contribution supported on the opposite one being at best a negligible tail. Therefore, we can approximate the state by its restriction (cutoff) to its dominating orientation and subsequently split the family on considered semiclassical states onto those peaked about positive and negative $k$ respectively. 

Using the volume operator as defined in \eqref{eq:V-gen} requires a knowledge on the action of the projection operators $P^{\pm}$ on the basis states \eqref{eq:WDW-basis-b}. That action has been determined in Appendix~\ref{app:WDW-b} and after applying an analogous to \eqref{eq:Gamma-dec} splitting of $\Gamma(ik)$ onto modulus and phase can be written as
\begin{equation}\label{eq:WDW-P1}
  [\mathcal{F}(P^{\pm}\ub{e}_k)](b) = \frac{\Gamma(ik)}{\sqrt{2}\pi} e^{\mp\sgn(b)k\frac{\pi}{2}} 
  e^{-ik\ln\left|\frac{b}{2}\right|} =: \frac{e^{\mp\sgn(b)k\frac{\pi}{2}}}{\sqrt{2\pi k\sinh(k\pi)}}
  e^{i\varphi(k)} e^{-ik\ln|b|} \ .
\end{equation}
For large $|k|$ these projections to great accuracy are projections onto appropriate semiaxis of $k$
\begin{equation}
  [\mathcal{F}(P^{\pm}\ub{e}_k)](b) = \frac{\theta(\mp\sgn(b)k)}{\sqrt{\pi|k|}} \left( e^{i\varphi(k)} e^{-ik\ln|b|} + O(e^{-k\pi}) \right) \ ,
\end{equation}
though, unlike in LQC prescriptions, after projection a nontrivial phase shift by $\varphi(k)$ remains. This however corresponds to freedom of choice of the basis, thus we can choose to work with one rotated appropriately to compensate for this offset
\begin{equation}\label{eq:WDW-rot}
  \Psi(v,\phi) = \int_{\re} \rd k \tilde{\Psi}(k) e^{-i\varphi(k)} \ub{e}_k(v) \ ,
\end{equation}
which in turn will simplify \eqref{eq:WDW-P1}, eliminating the phase rotation term. Then upon setting $x:=\ln(b)$ we can adapt the definition \eqref{generalformofvolumoperator} introducing a pair of operators
\begin{equation}
  \tilde{V}^{\pm} := \mp iA f(x)\partial_x \ub{\mathcal{P}}^{+} 
  = \mp i \frac{\gamma\hbar\sqrt{\pi G}}{\sqrt{3}} \exp(x)\partial_x \ , \qquad 
  [\ub{\mathcal{P}}^{\pm} e^{-i\varphi(k)} \ub{e}_k](x) 
  := \frac{\theta(\mp k)}{\sqrt{\pi|k|}} e^{-ikx} \ ,
\end{equation}
of which $\tilde{V}^+$ will extract the part corresponding to $k<0$ (expanding in $v$) from $v>0$ and $\tilde{V}^-$ will extract part corresponding to $k>0$ (contracting in $v$) from $v<0$.
This pair will then replace $\tilde{V}^+$ in the evaluations of sec.~\ref{sec:semiclass-eom}. Note, that this will change the relation between the auxiliary and physical variables \eqref{eq:phys-var} as well as the central moments \eqref{eq:phys-G}, as we want to keep $\hat{P}_{\phi}$ non-negative. Thus
\begin{equation}\label{eq:var-WDW}
  \hat{p}_{\phi} = \hbar\beta|k| = -\sigma\beta k \ , \qquad
  \hat{\phi}_{0} = -i\sigma\beta^{-1} \partial_k \ , \qquad
  G^{a,b} = (-\sigma)^{a+b} \hbar^a \beta^{a-b} \tilde{G}^{a,b} \ ,
\end{equation}
where $\sigma=+1$ for the ever-expanding (peaked about $k_0<0$) and $\sigma=-1$ for the ever-contracting (peaked about $k_0>0$) semiclassical states.

Applying the above, we arrive to the following results regarding the volume trajectory and the (squared) variance (which now can be presented with all the orders of semiclassical corrections)
\begin{subequations}\label{eq:traj-WDW-both}\begin{align}
  \begin{split}\label{finalformforexpVwdw}
    \braket{\hat{V}_{\phi}} 
    &= \frac{\gamma\sqrt{\pi G\Delta }}{\sqrt{3} }\exp(\sigma\beta(\phi-\phi_0)) \left[ p_{\phi} + \sum_{i=1}^{\infty}\frac{(-\sigma\beta)^{i}}{i!}\Big(p_{\phi} G^{0,i}+G^{1,i}\Big) \right] + O(p_{\phi}^{-1}) \ ,
  \end{split} \\
  \begin{split}\label{stdgeo}
    \sigma^2(V_{\phi}) 
    &= \frac{\gamma^2\pi G\Delta }{3}\exp(\sigma\beta(\phi-\phi_0))\sum_{i=1}^{\infty}\Big[p_{\phi}^2\big(
    \sum_{a=0}^i\frac{(-\sigma\beta)^{i}}{(i-a)!a!}G^{0,i}-\sum_{i'=0}^{\infty}\frac{(-\sigma\beta)^{i}}{i!}\frac{(-\sigma\beta)^{i'}}{i'!}G^{0,i}G^{0,i'}\big) \\
    &+ 2p_{\phi}\big(
    \sum_{a=0}^i\frac{(-\sigma\beta)^{i}}{(i-a)!a!}G^{1,i}
    -\sum_{i'=0}^{\infty}\frac{(-\sigma\beta)^{i}}{i!}\frac{(-\sigma\beta)^{i'}}{i'!}G^{0,i}G^{1,i'}\big) \\
    &+ \big(
    \sum_{a=0}^i\frac{(-\sigma\beta)^{i}}{(i-a)!a!}G^{2,i}-\sum_{i'=0}^{\infty}\frac{(-\sigma\beta)^{i}}{i!}\frac{(-\sigma\beta)^{i'-1}}{i'!}G^{1,i}G^{1,i'}\big)\Big] 
    + O(p_{\phi}^0) \ .
  \end{split}
\end{align}\end{subequations}
At this point several comments are in order:
\begin{enumerate}
  \item While it is known, that for the model under consideration the quantum trajectory coincides with the classical one, in the above equation we see an explicit dependence of $V$ on $\hbar$. Note however, that (due to exponential form of the trajectory) rescaling of $V$ is equivalent to a shift in $\phi_0$ which in turn is equivalent to certain $k$-dependent rotation of the basis vectors. Thus, that dependence is spurious -- tied to our choice of the basis.
  \item The calculations of $\braket{\hat{V}_{\phi}}$, $\sigma^2(V_{\phi})$ can be (much easier) performed 
  directly in $v$-representation (using the original basis \eqref{eq:WDW-basis}), where $V$ is a multiplication operator. In terms of the variables and central moments defined analogously to \eqref{eq:var-WDW} they will take the form \eqref{eq:WDW-VdV-v}\footnote{In order to bring the expression to the form similar to \eqref{eq:traj-WDW-both} we rescaled the global constant, compensating the rescaling by a shift inside the exponent.} (see Appendix~\ref{sec:WDW-traj-v} for the calculations)  
  \begin{subequations}\label{eq:WDW-VdV-v-final}\begin{align}
    \braket{\hat{V}_{\phi}} 
    &= \frac{\gamma\sqrt{\pi G\Delta }}{\sqrt{3}} p_{\phi} \exp\left(\sigma\beta(\phi -\phi_0)-\ln\left(\frac{p_{\phi}}{\beta\hbar}\right)\right)\sum_{l=0}^\infty \frac{(-\sigma\beta)^l}{l!} G^{0,l}   \\
    \sigma^2(\hat{V}_{\phi})
    &= \frac{\gamma^2\pi G\Delta }{3} p_{\phi}^2 \exp\left(2\sigma\beta(\phi -\phi_0) -2\ln\left(\frac{p_{\phi}}{\beta\hbar}\right) \right) \sum_{l=0}^\infty \frac{(-\sigma\beta)^l}{l!} \left(
    2^l G^{0,l} - \sum_{n=0}^l \binom{l}{n} G^{0,n} G^{0,l-n}
    \right) \ .
  \end{align}\end{subequations}
  which: $(i)$ no longer contains corrections proportional to the negative powers of $p_{\phi}$, and $(ii)$ differs from \eqref{eq:traj-WDW-both} being in fact much simpler. The reason for this difference is a discrepancy in the expectation values and moments involving the operator $i\partial_k$ in both approaches, as the nontrivial phase rotation \eqref{eq:WDW-rot} changes their values.
\end{enumerate}
As we can see, with very few minor corrections, the proposed computation technique works very well also when applied to the Wheeler-DeWitt description. 

Let us now focus on the last prescription listed explicitly in our work -- the one following from the strict application of the Thiemann regularization algorithm.

\subsubsection{Strict Thiemann regularization}

The model resulting from applying this regularization prescription differs in a few crucial points from the other ones. First of all, while after transforming to $b$ representation (with transformation rules the same as for mainstream LQC) it is still possible to bring the evolution operator to Klein-Gordon form, the whole Hamiltonian constraint changes the signature depending on the value of the coordinate $x(b)$, namely
\begin{equation}
  \Theta = -12\pi G\, \sgn(|x|-\pi/2) \partial_x^2 \ ,
\end{equation}
where $x(b)$ is known explicitly in analytic form (see \cite{Assanioussi:2019iye} for the details. Furthermore, $\Theta$ admits a $1$-dimensional family of selfadjoint extensions distinguished by glueing conditions at $x=\pm\pi/2$. The Hilbert space is spanned by energy eigenstates which have a form of standing (reflected) plane wave for $|x|>\pi/2$ and are exponentially suppressed for $x\in[-\pi/2,\pi/2]$ which is a classically forbidden region. Spectrum of the positive part of $\Theta$ is continuous (in fact ${\rm Sp}(|\Theta|)=\re^+$ and on the subspace of states symmetric in $v$ (which is equivalent to symmetry in $x$) nondegenerate. Therefore, physical states have in the energy representation same structure as in other prescriptions, the only element that differs is the exact form of energy eigenstates. One could thus expect, that the technique of extracting the quantum trajectories presented in this paper should work also for this model. Because of the standing wave form of the eigenstates, even the approximation of the volume operator proposed in \eqref{generalformofvolumoperator} seems to be possible. The rotating components of the energy eigenstates have nontrivial $k$ and extension dependent phase shifts, thus in reexpressing the 
operators, like $\tilde{V}^+_{\phi}$ in $k$-representation one would need to account for them. This however is just a technical problem. It follows then, that by applying our procedure one would arrive to a definite expression for the volume trajectory and dispersion. It is even quite possible that in $0$th order the result would agree with the heuristic effective dynamics. Does it mean, that our calculation procedure works also in this case? 

Unfortunately, the answer to this question is in the negative. The reason for that lies in the assumptions, that allowed us to derive the expressions \eqref{finalformforexpV} and \eqref{eq:sigma2-final}. In particular, the implementation of the Taylor expansion\footnote{Here the expansion would need to be performed about some $x_0$ within classically allowed region instead of $x=0$} in \eqref{expectationV} assumes finiteness of the expectation values. However, the action of the operator $\hat{V}_{\phi}$ actually leads outside the physical Hilbert space. There is no semiclassical sector where the states of spectral profiles $\Psi(k)$ belonging to Schwartz space would have a finite volume. This problem has already been discussed in \cite{Assanioussi:2019iye} and in more detail in the model admitting positive cosmological constant \cite{Pawlowski:2011zf} which shares this feature. There, the problem has been circumvented by using instead of $V$ a \emph{regularized volume} corresponding to some globally bounded function of $V$. 
A suitable generalization of the computation procedure implementing such regularized observables, if possible, is a matter for potential future research.

One final issue that needs discussing is an apparent discrepancy of our results regarding trajectories with the result of \cite{Kaminski:2019tqo}. There, it was reported, that for both mainstream LQC prescription and Wheeler-DeWitt analog of the model studied here, the states disperse immediately once only the positive energy sector is included.  This problem is encountered already in the textbook example of the Klain-Gordon equation, once we consider $\exp(x)$ as a position observable of interest (in which case the large semiclassical sector is known to exist). One of the possible solutions to the problem is to include both positive and negative energy sectors. Then a small negative energy tail can regularize the singularity. One can apply the very same approach here. Then, since we consider states peaked sharply at large $k$ (in the positive energy sector) such tail, besides providing the necessary regularization, would be much smaller, than the terms already neglected within the approximation taken. Thus, in the method devised here, we sidestep the problem indicated \cite{Kaminski:2019tqo} without apparent consequences on the dynamical predictions.

\section{Conclusions}
\label{sec:conclusions}

In this paper, we explored the freedom in the Thiemann regularization procedure in Loop Quantum Gravity in the context of Loop Quantum Cosmology. Using an example of a flat Friedmann-Lemaitre-Roberton-Walker universe with a massless scalar field we reviewed and compared the three regularization prescriptions substantially discussed in the literature so far. Two of them have been analysed in detail in the literature on the genuine quantum level. Since for the third one (here denoted as Yang-Ding-Ma (YDM) prescription) the studies so far focused mainly on quantum kinematics and phenomenology of the dynamics, we performed a rigorous analysis of this model. This brought that prescription to the same footing as the other considered.

In more detail, in the analysis of YDM prescription, we reexamined the properties of the so-called evolution operator (in the model playing the role of a square of the Hamiltonian), in particular establishing its essential selfadjointness (through an analysis of its deficiency subspaces) and subsequently identifying its spectrum. Further construction of an equivalent of the ``energy'' eigenbasis allowed to explicitly build a physical Hilbert space via group averaging -- a technique already used in the literature for the other prescriptions. Despite being (in volume representation) a difference operator of the 4th order the evolution operator has shown properties very similar to the 2nd order one in the mainstream LQC: its spectrum is (within the probed superselection sector) nondegenerate and the eigenfunctions forming the basis of the geometry (or gravitational) Hilbert space have a form of standing waves reflected certain ``distance'' from the classical singularity and in large volume limit converge to a standing wave of the geometrodynamical (Wheeler-DeWitt) analog of studied model. This ensures the existence of a large semiclassical sector within the model at least in the distant future or past. These properties allowed in turn to apply here a convenient set of partial observables already used in LQC in order to probe the quantum dynamics.

In order to extract physical predictions we analysed the trajectory (and variances) of the volume of the (portion of the) Universe at a given value of the scalar field (working as internal clock). Unlike in other prescriptions though, instead of resorting to numerical methods (here a bit more difficult to implement due to specific properties of the YDM prescription) here, we devised a relatively simple (though sometimes computationally tedious) analytical method based on evaluating the desired expectation values on the variable classically corresponding to the canonical momentum of the volume) and transforming the result to a form of an explicit function of the scalar field ``time'' and a series of the so-called central (Hamburger) moments built off a canonical pair of Dirac observables (both constants of motion): the scalar field momentum $\hat{p}_{\phi}$ and a certain ``scalar field offset'' $\hat{\phi}_0$.  That method allowed in particular to find the explicit form of the trajectory in volume $V(\phi)=\braket{\hat{V}_{\phi}}$ (and its variance) of Universe sharply peaked at large value of scalar field momentum up to an arbitrary order of quantum corrections.  

Application of this method to YDM prescription produced an expression for $V(\phi)$ and its variance out of which we wrote down in an explicit closed form the terms up to 2nd order (leaving remaining ones as expressed through $n$th order derivatives of a certain function). The leading ($0$th order) term was shown to agree with the already known effective trajectory derived by phenomenological methods, thus confirming the conclusions drawn from the latter: the existence of two large epochs of contracting and expanding Universe (evolving to great accuracy as predicted by General Relativity) connected by a single quantum bounce, happening at certain critical energy density of the scalar field, of which value differs from that of mainstream LQC but remains of Planck order. 

The leading ($2$nd) order of corrections is a combination of terms proportional to the variance in the offset $\phi_o$ and its correlation with $p_{\phi}$ (with the first one dominating the correction). The very same terms enter the leading order of the variance. The expression for the latter has been used to probe the possible semiclassicality loss between the distant future and past. The difference of the asymptotic (future and past) relative variances is to the leading order bounded by the correlation $p_{\phi}$--$\phi_0$, which provides a severe limitation on the growth of the variances (at distant time) as that correlation is is in turn bounded by the variances of $\phi_0$ and $p_{\phi}$. 

In order to check the robustness of the method, it was applied also to the mainstream LQC and (after minor adaptation) to the Wheeler-DeWitt analog. There, as the models are a bit simpler, it was possible to write in a compact form explicit closed formulas for all orders of corrections to $V(\phi)$. Again, the leading order terms agreed with those found via phenomenological methods and the corrections were governed by the $2$nd and higher order central moments of $p_{\phi}$ and $\phi_0$. 

However, to show also the limitations of the technique a discussion of an application of it to a model based on strict Thiemann regularization was also included. While its blind application would yield definite results, they will not be reliable due to lack (in the model) of the large sector, that is semiclassical in variables used by the method. This emphasizes the need to check the structure of the physical Hilbert space and the action of observables of interests in it before fully trusting semiclassical methods.

One of the most prominent pros of the implemented method is its robustness. It allows to control the dynamics of the quantum state without resorting to numerical methods for a wide variety of models, either in LQC or geometrodynamics. Its strength is particularly visible in YDM model, where using the standard numerical methods of LQC to probe the genuine quantum mechanics is difficult due to instabilities.
Presented alternative is directly applicable to other isotropic models, like the ones with cosmological constant (as the volume still has the form \eqref{eq:V-gen} there), or models with different matter clocks, as well as the nonisotropic ones (like Bianchi I, see in particular \cite{Martin-Benito:2011fdk}). It may also constitute a good point of departure for generalization to the inhomogeneous ones.

However, one still needs to remember that, despite including contributions from all orders of quantum corrections the method devised here is not exact. In order to deal with the commutators with $\sqrt{k}$ we introduced an approximation, neglecting in $V(\phi)$ and $\sigma(V_{\phi})$ the terms scaling with negative powers of $p_{\phi}$. While these terms could be in principle included, this would make all the expressions much more complicated. Furthermore, in order to be able to perform the expansion we needed to strengthen the restrictions on the semiclassical states we are working with, requiring that the states under consideration have finite expectation values of $\braket{\hat{p}_{\phi}^n}$. This assumption however does not appear restrictive when we consider the states interesting from a physical point of view -- sharply peaked at large values of scalar field momentum (or other equivalent of the energy in cases of using different matter fields as clocks). The strengthened restriction is however tied to specific calculation in rewriting specific expectation values in terms of those of Weyl-ordered operators, thus may not be a necessary condition for the approach to work.

\begin{acknowledgments}
    This work was supported in part by the Polish National Center for Science (Narodowe Centrum Nauki -- NCN) grant OPUS 2020/37/B/ST2/03604.
\end{acknowledgments}

\appendix

\section{Wheeler-DeWitt analog of the model}
\label{sec:WDW}

Here we briefly recall the main features of a quantum cosmological model constructed via treating the classical model of the flat FRLW universe with massless scalar field via methods of geometrodynamics -- the so-called Wheeler-DeWitt (WDW) model \cite{Kiefer:1988tr}. Such model has been extensively discussed in the literature (for the details relevant for our work see for example \cite{Ashtekar:2006uz}). The starting point is the same as in case of LQC quantization -- the canonical formulation of GR reduced to the isotropic setting, as in sec.~\ref{sec:class_kin}. One can choose to use the triad formalism, with partial gauge fixing and variables defined as in \eqref{eq:EA-iso}, though it is not necessary -- one can start with the ADM formalism, relating the global coefficients $(v,b)$ with the scale factor and the Hubble parameter as in \eqref{eq:vb-def}. Both approaches are mathematically equivalent. 

Subsequently one applies to the pair of variables $(v,b)$ the standard textbook Schr\"odinger quantization procedure. As a consequence, the family of holonomies along straight lines is now continuous, allowing, in particular, to take the limit in \eqref{equA}. As a consequence, the gravitational component of the Hamiltonian constraint reduced to just a simple function of $v$ and $b$
\begin{equation}
  H = -\frac{3\pi G\hbar^2}{2\alpha} v b^2 \ ,
\end{equation}
where $\alpha$ is the constant defined in \eqref{eq:vb-def}. Coupling to it the massless scalar field (described by a standard canonical pair $(\phi,p_\phi)$) and choosing the lapse $N=2V$ we arrive to a total Hamiltonian constraint of the form
\begin{equation}
  C_{{\rm tot}} = p_{\phi}^2 - 3\pi G\hbar^2 v^2 b^2 \ .
\end{equation}
Since the process of quantization involves standard quantum mechanics procedures, the kinematical Hilbert space of the system will be just the product of two Lebesgue spaces
\begin{equation}
  \ub{\Hil}_{{\rm kin}} = L^2(\re,\rd v) \otimes L^2(\re, \rd \phi) \ .
\end{equation}
Upon promoting basic variables to operators this constraint takes the form of a Klain-Gordon equation
\begin{equation}
  \hat{\ub{C}}_{{\rm tot}} = -\mathbb{I}_{{\rm gr}}\otimes\hbar^2\partial_{\phi}^2
  - \ub{\Theta}\otimes\mathbb{I}_{\phi} \ , 
  \qquad
  \ub{\Theta} = -12\pi G\hbar^2 (\sqrt{|v|}\partial_v\sqrt{|v|})^2 \ .
\end{equation}
By applying to it the group averaging procedure and subsequently choosing the superselection sectors corresponding to states of positive energy\footnote{This is the standard choice for the Klain-Gordon equation. Here we consider the scalar field as an evolution parameter -- an internal clock. In such case, the momentum $p_{\phi}$ plays the role of 'energy' of the system.} and symmetric in $v$ one arrives to the following form of physical states
\begin{equation}\label{eq:WDW-state}
  \Psi(v,\phi) = \int_{\re} \rd k \tilde{\Psi}(k) \ub{e}_k(v) e^{i\omega(k)\phi} \ ,
\end{equation}
where the energy eigenstates $e_k$ take the form\footnote{The normalization factor differs from that in \cite{Assanioussi:2019iye} since here we integrate over both orientations of $v$.}
\begin{equation}
  \ub{e}_k(v) = \frac{1}{\sqrt{4\pi|v|}} e^{ik\ln|v|}
  \label{eq:WDW-basis}
\end{equation}
and the dispersion relation $\omega(k)$ is
\begin{equation}\label{eq:WDW-disp}
  \omega(k) = \sqrt{12\pi G}|k| \ .
\end{equation}
This is an equivalent of a decomposition of a solution to the K-G equation onto plane waves, where the eigenstates for $k<0$ / $k>0$ correspond to the outgoing (expanding) / incoming (collapsing) waves respectively.

\subsection{Hubble rate representation}
\label{app:WDW-b}

For the purpose of comparizon with the LQC model studied in this article it is convenient to also recall the form of the physical states, inner products and observables expressed in momentum $b$ representation. That form has been considered in detail already in \cite{Ashtekar:2007em} (with further corrections in \cite{Bodendorfer:2018zyr}). In what follows we will use the conventions and notation from Appendix C of \cite{Assanioussi:2019iye} further providing some minor corrections.

Unlike in LQC, here $\hat{b}$ is a well defined operator and we can pick a basis formed off its generalized eigenstates. One can define the  transformation (and its inverse) from the $v$ basis to it as 
\begin{equation}
  \label{eq:vb-trans-WDW}
    [\ub{\mathcal{F}}\psi](b) 
    = \frac{1}{2\sqrt{\pi}} \int_{\re} \frac{\rd v}{\sqrt{|v|}} \psi(v) e^{i\frac{vb}{2}} \ ,
    \qquad
    [\ub{\mathcal{F}}^{-1}\psi](v) 
    = \frac{\sqrt{|v|}}{2\sqrt{\pi}} \int_{\re} \rd b \psi(b) e^{-i\frac{vb}{2}} \ .
\end{equation}
The basic operators take in the new representation the form
\begin{equation}
  \hat{v} = -2i\partial_b \ , 
  \qquad
  \sqrt{|\hat{v}|}\hat{b}\frac{1}{\sqrt{|\hat{v}|}} = 2ib\mathbb{I} \ ,
\end{equation}
however, due to presence of $|v|$ in the transformation, the inner product of $\Hil_{\rm gr}$ cannot be expressed locally, however one can introduce a projection onto orthogonal subspaces $\Hil^{\pm}$ supported on the positive/negative $v$
\begin{equation}
    P^{\pm}:\ub{\Hil}_{\rm gr}\to\ub{\Hil}^{\pm}\subset \ub{\Hil}_{\rm gr} \ , 
    \qquad 
    [P^{\pm}\psi](v) = \psi(v)\theta(\pm v) \ ,
\end{equation}
where $\theta$ is the Heaviside step function. On these subspaces the induced inner product can be expressed in a local form
\begin{equation}
  \langle \psi|\chi \rangle = \langle \psi|P^+\chi \rangle_+ + \langle \psi|P^-\chi \rangle_- \ ,
  \qquad 
  \langle \psi|\chi\rangle_{\pm} = \mp 2i\int_{\re} \rd b \bar{\psi}(b)\partial_b\chi(b) \ .
\end{equation}
The volume operator can now also be expressed in a simple way
\begin{equation}
  \hat{V} = \alpha|\hat{v}| = -2i\alpha\partial_b(P^+-P^-) \ .
\end{equation}
While the projection operators $P^{\pm}$ cannot be written down as easily, their action on the basis elements can be evaluated. For the details, we refer the reader to the Appendix C of \cite{Assanioussi:2019iye}. Here we present the result, correcting some minor sign discrepancies
\begin{equation}\label{eq:Proj_ube}
  [\ub{\mathcal{F}}(P^{\pm}\ub{e}_k](b) 
  = \frac{\sqrt{2}}{4\pi}\int_{\re^\pm} \frac{\rd v}{|v|} e^{ik\ln|v|} e^{i\frac{vb}{2}} 
  = \frac{\Gamma(ik)}{\sqrt{2}\pi} e^{\mp\sgn(b)k\frac{\pi}{2}} e^{-ik\ln\left|\frac{b}{2}\right|} \ .
\end{equation}
Note, that one sign of $b$ is always suppressed, with the suppression factor exponential in $|k|$. Thus for large $|k|$ the projected base function can be approximated as
\begin{equation}\label{eq:WDW_basis_proj}
  \forall |k|\gg 1 \ , \quad 
  [\ub{\mathcal{F}}(P^{\pm}\ub{e}_k](b) 
  \approx \theta(\mp k b) \frac{\Gamma(ik)}{2\sqrt{2}\pi} e^{|k|\frac{\pi}{2}} e^{-ik\ln\left|\frac{b}{2}\right|} 
  \ .
\end{equation}
Summing up the two terms in \eqref{eq:Proj_ube} we finally write the full basis element\footnote{Note the change in the formula with respect to (C16) of \cite{Assanioussi:2019iye} resulting from sign corrections in \eqref{eq:Proj_ube}. This change does not however affect the large $k$ behaviour of $\ub{e}_k$ thus the approximations taken for the normalization of $e_{\beta,k}$ in \cite{Assanioussi:2019iye} remain valid.}
\begin{equation}\label{eq:WDW-basis-b}
  [\ub{\mathcal{F}}e_k](b) 
  = \frac{\Gamma(ik)}{\sqrt{2}\pi} \cosh\left(\frac{\pi}{2}k\right) e^{-ik\ln\left|\frac{b}{2}\right|}
  =: \ub{B}_k e^{-ik\ln\left|\frac{b}{2}\right|} \ .
\end{equation}
Using known mathematical formula $\Gamma(z)\Gamma(-z) = -\pi/(z\sin(\pi z))$ one can evaluate the normalization factor explicitly
\begin{equation}\label{eq:WDW-x-norm}
  |B_k| = \frac{\cosh(k\pi/2)}{\sqrt{2\pi k\sinh(k\pi)}} 
  = \frac{1}{2\sqrt{\pi|k|}} \left( 1 + O\left(\frac{e^{-|k|\pi/2}}{\sqrt{|k|}}\right) \right) \ .
\end{equation}

\section{Wheeler-DeWitt limit of LQC model}
\label{sec:WDW-limit}

In case of the prescriptions already analyzed in the literature in detail: mainstream LQC and strict Thiemann regularization (presented in sec.~\ref{sec:mainstream} and \ref{sec:Thiemann} respectively),
it was possible to identify a well defined geometrodynamic (WDW) limit of each model. Namely, each LQC energy eigenfunction asymptotically approached some combination of its analog in WDW approach. This feature allowed to look at the loop quantum geometry modifications as a process of ``scattering'' of WDW universe, while on technical level it made possible to evaluate (fix) the normalization of the LQC energy eigenfunctions (see Appendix A.2 of \cite{Kaminski:2010yz} and Appendix D.3 of \cite{Assanioussi:2019iye}). It is thus expected that the model following from Yiang-Ding-Ma prescription shares the same features. We will verify this expectation below.

Our point of departure is the (determined up to a normalization factor) analytic form of the symmetric energy eigenfunctions (corresponding to the eigenvalue $\omega^2$) in $b$-representation 
\begin{equation}
  \psi_k(b) = N_k\cos(kx(b)) \ , \qquad \omega = \beta k \ ,
\end{equation}
where $x(b)$ is defined via \eqref{eq:x}.

Since in $b$-representation the inner product has no simple form (unless the states are projected onto particular orientation of $v$, which is difficult to control in $b$-representation itself) it is beneficial to convert the above eigenfuction to $v$-representation. This is provided via the transformation \eqref{eq:vb-trans-inv}
\begin{equation}
  \mathcal{F}^{-1}\psi(v) = \frac{\sqrt{|v|}}{2\pi} N_k \left( \xi_k(v) + \xi_{-k}(v) \right) \ , 
  \quad 
  \xi_k(v) := \int_0^{2\pi}\rd b e^{ikx(b)} e^{-\frac{ivb}{2}} \ .
  \label{eq:e-inv-app}
\end{equation}
Due to symmetry of $\psi$ it is enough to evaluate the above integral for $v>0$ only. To do so, we analytically extend the integrand to the complex plane and consider a contour integral over the closed path
\begin{equation}\label{eq:xi_intS}
  S_k(v) = \lim_{\epsilon\to 0,R\to\infty} \oint_{C(\epsilon,R)} \rd b e^{ikx(b)} e^{-\frac{vb}{2}} \ ,
\end{equation}
where the contour $C(\epsilon,R)$ is defined as follows (see Fig.~\ref{integration contours-LQC})
\begin{figure}
    \centering
    \includegraphics[width=0.45\textwidth]{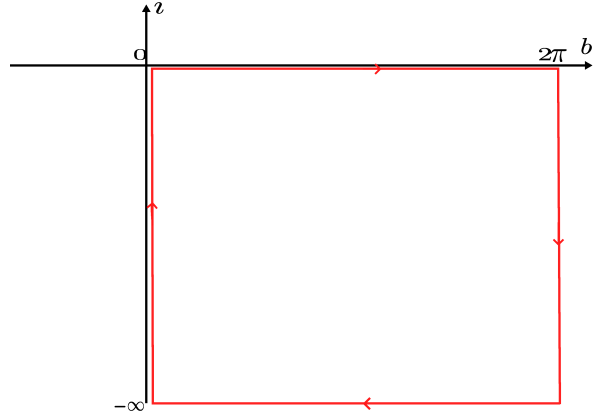}
    \caption{integration contours}
    \label{integration contours-LQC}
\end{figure}
\begin{equation}
  C(\epsilon,R) := [\epsilon,2\pi-\epsilon]\cup[2\pi-\epsilon,2\pi-\epsilon-iR]\cup[2\pi-\epsilon-iR,\epsilon-iR]\cup[-iR,\epsilon]
\end{equation}
By direct inspection the integrand of \eqref{eq:xi_intS} is regular inside the contour. Furthermore, the integral over $[2\pi-\epsilon-iR,\epsilon-iR]$ vanishes in the limit $R\to\infty$. Thus, by Cauchy theorem
\begin{equation}\begin{split}
    \xi_k(v) 
    &= -i\int_0^{\infty}d\lambda e^{-\frac{v\lambda}{2}}e^{ikx(-i\lambda)} + i\int_0^{\infty}d\lambda e^{-\frac{v(\lambda-2i\pi)}{2}}e^{ikx(2\pi-i\lambda)}
    \\
    &= -i\int_0^{\infty}d\lambda e^{-\frac{v\lambda}{2}}e^{ikx(-i\lambda)} + i\int_0^{\infty}d\lambda e^{-\frac{v\lambda}{2}}e^{-ikx(i\lambda)}
    = -\int_0^{\infty}d\lambda e^{-\frac{v\lambda}{2}}(I_k+I_k^{\dagger}) \ ,
  \label{onebranchintegratedovercountur}
\end{split}\end{equation}
where $I_k:=ie^{ikx(-i\lambda)}$ and in the 2nd line we used the fact, that under our selection of the superselection sector we have $v\in2\mathbb{Z}^+$ and the antisymmetry $x(2\pi-i\lambda)=-x(i\lambda)$.

Next, we separate out the singular terms of the integrand (singular at $b\in\{0,2\pi\}$) by defining the function
\begin{equation}
  \sigma(\lambda) := x(-i\lambda)-\ln{(-i\lambda)}+\ln{(2\pi+i\lambda)}
\end{equation}
and rewriting $I_k$ as
\begin{equation}\begin{split}
    I_k 
    &= i(e^{ik\sigma(-i\lambda)}e^{ik\ln{(-i\lambda)}}e^{-ik\ln{(2\pi+i\lambda)}} 
    -a_+e^{ik\ln{(-i\lambda)}}-a_-e^{-ik\ln{(2\pi-i\lambda)}}) 
    \\ 
    &+ ia_+e^{ik\ln{(-i\lambda)}} + ia_-e^{-ik\ln{(2\pi-i\lambda)}} \ ,
\end{split}\end{equation}
where constants $a_{\pm}$ are determined by the requirement, that the first term is free from the logarithmic singularity at $b\in\{0,2\pi\}$ and equal
\begin{equation}\label{eq:sigma0-def}
  a_+ = \exp(ik\sigma_0) \ , \qquad 
  a_- = \exp(-ik\sigma_0) = \bar{a}_+ \ , \qquad 
  \sigma_0 := \sigma(0)-\ln(2\pi) = \ln{(\frac{1}{4}\sqrt{1+\gamma^2})} \ .
\end{equation}
Under that choice, it is convenient to rewrite the first term of $I_k$ using well-known identity $e^a-a^b=2ie^{\frac{a+b}{2}}\sin{\frac{a-b}{2}}$ so that
\begin{equation}\begin{split}
    I_k &= -2e^{ik\ln{(-i\lambda)}}e^{\frac{ik}{2}(\sigma - \ln(2\pi+i\lambda)+\sigma_0)}\sin\left(\frac{k}{2}(\sigma - \ln{(2\pi+i\lambda)-\sigma_0})\right)+ia_+e^{ik\ln{(-i\lambda)}} 
    \\ 
    &=: \delta I_k + \ub{I}_k \ ,
\end{split}\end{equation}
where $\delta I_k$ and $\ub{I}_k$ denote the 1st and the 2nd term respectively. The latter is easy to integrate, with the result being proportional to the WDW analog model basis function \eqref{eq:WDW-basis}
\begin{equation}\begin{split}
  \int_0^{\infty}\rd\lambda e^{-\frac{v\lambda}{2}} \ub{I}_k 
  &= ie^{ik\sigma_0} \int_0^{\infty}\rd\lambda e^{-\frac{v\lambda}{2}} e^{ik\ln(-i\lambda)}
  = -\frac{2k}{|v|} e^{ik(\sigma_0+\ln(2))} e^{\frac{k\pi}{2}} \Gamma(ik) e^{-ik\ln|v|} \\
  &= -4 \sqrt{\frac{\pi}{|v|}} e^{ik(\sigma_0+\ln(2))} e^{\frac{k\pi}{2}} \Gamma(ik)\, \ub{e}_{-k}(v) \ .
  \label{eq:ubik-int}
\end{split}\end{equation}
In order to estimate the integral of $\delta I_k$ we split it onto the regular and singular terms
\begin{equation}
  \delta I_k(\lambda) = f(\lambda)g(\lambda) \ ,    
\end{equation}
where
\begin{equation}
  f(\lambda) = e^{ik\ln{(-i\lambda)}} \ , \qquad 
  g(\lambda) = e^{\frac{ik}{2}(\sigma(\lambda) - \ln{(2\pi+i\lambda)+\sigma_0})}
  \sin\left(\frac{k}{2}(\sigma(\lambda) - \ln{(2\pi+i\lambda)-\sigma_0 })\right) \ .
\end{equation}
By Taylor expanding $g$ about $\lambda=0$ we get 
\begin{equation}
  g(\lambda) = \frac{k(3\gamma^2-1)}{96}e^{ik\ln{\sigma_0}}\lambda^2+O(\lambda^3) \ .
  \label{eq:g-exp}
\end{equation}
This allows to bound the modulus of the integral of $\delta I_k$ as
\begin{equation}
  \left|\int_0^{\infty}\rd\lambda e^{-\frac{v\lambda}{2}} \delta I_k(\lambda) \right|
  \leq \int_0^{\infty}\rd\lambda e^{-\frac{v\lambda}{2}} |f(\lambda)| |g(\lambda)| 
  = \frac{k(3\gamma^2-1)}{6} e^{\frac{k\pi}{2}} |v|^{-3} + O(v^{-4}) \ ,
\end{equation}
where to get the last expression we applied the expansion \eqref{eq:g-exp} and the equality $|f(\lambda)|=\exp(k\pi/2)$. 

Plugging the above estimate and \eqref{eq:ubik-int} back into \eqref{onebranchintegratedovercountur} 
we arrive to the following result
\begin{equation}\label{eq:xi-form}
  \xi_k(v) = 4\sqrt{\frac{\pi}{|v|}} k e^{\frac{k\pi}{2}} 
  \left[\Gamma(-ik) e^{-ik(\sigma_0+\ln(2))} \ub{e}_{k}(v) + \Gamma(ik) e^{ik(\sigma_0+\ln(2))} \ub{e}_{-k}(v) \right]
  + O(v^{-2})
  \ ,
\end{equation}
subsequently, inserting the above into \eqref{eq:e-inv-app} we finally get
\begin{equation}
  \psi_k(v) = N_k \frac{4}{\sqrt{\pi}} k \sinh\left(\frac{k\pi}{2}\right) 
    \left[ \Gamma(-ik) e^{-ik(\sigma_0+\ln(2))} \ub{e}_k(v) + \Gamma(ik) e^{ik(\sigma_0+\ln(2))} \ub{e}_{-k}(v) \right] 
    + O(|v|^{-3/2})\ ,
  \label{eq:e-LQC-as}
\end{equation}
which by symmetry of $\psi(v)$ is valid also for $v<0$.

Consider now a scalar product of two eigenfunctions. By repeating the reasoning already discussed in Appendix D.3 of \cite{Assanioussi:2019iye}, that is by noting that:
\begin{enumerate}
  \item since the leading terms in \eqref{eq:e-LQC-as} decay like $|v|^{-1/2}$, while the corrections decay like $|v|^{-3/2}$ all the cross terms and products of corrections in the inner product will have finite contribution. and
  \item within the domain of $|v|>1$ the sum in the LQC inner product can be approximated by an integral, with the approximation error being again finite
\end{enumerate}
we can write the scalar product as
\begin{equation}\label{eq:ip-app}
  \langle \psi_k | \psi_{k'} \rangle = \frac{16}{\pi} k^2 N_k^2 \sinh^2\left(\frac{k\pi}{2}\right) |\Gamma(ik)|^2 \delta(k-k') + f(k,k')\ , 
\end{equation}
where $f(k,k')$ can a-priori be singular at $k=k'$. However, the orthogonality of eigenspaces for $k\neq k'$ implies $f(k,k')=0$. 
Consequently, applying the identity already used in Appendix~\ref{app:WDW-b} $|\Gamma(ik)|^2=\pi/(k\sinh(k\pi))$ we can determine $N_k$ corresponding to normalized eigenfunction
(denoted as $e_k(v)$)
\begin{equation}\label{eq:LQC-normalization}
  N_k = \frac{1}{4\sqrt{k}} \frac{\sqrt{\sinh(k\pi)}}{\sinh(k\pi/2)} = \frac{1}{2\sqrt{2k}} \left( 1 + O(e^{-k\pi/2}) \right) \ .
\end{equation}

Let us go back to the function $\xi_k(v)$. For $v>0$ it is given by eq.\eqref{eq:xi-form}. In order to determine it for negative $v$ we apply the symmetry following from \eqref{eq:vb-trans-inv} $\xi_k(-v) = \overline{\xi_{-k}(v)}$ which gives
\begin{equation}\label{eq:xi-form-neg}
  \forall v<0 \  
  \xi_k(v) = -4\sqrt{\frac{\pi}{|v|}} k e^{-\frac{k\pi}{2}} 
  \left[\Gamma(-ik) e^{-ik(\sigma_0+\ln(2))} \ub{e}_{k}(v) + \Gamma(ik) e^{ik(\sigma_0+\ln(2))} \ub{e}_{-k}(v) \right]
  + O(|v|^{-2}) \ . 
\end{equation}
Together with \eqref{eq:xi-form} this result allows to write down the (defined in sec.~\eqref{thetaB}) projections $P^{\pm}$ of $e_k(v)$ onto positive/negative $v$ semilattices as combinations of $\xi_{\pm k}$
\begin{equation}\label{eq:proj_ev}
  [P^{\pm}e_k](v) = \frac{1}{2\pi} \sqrt{\frac{|v|}{2k\sinh(k\pi)}} 
  \left[ e^{\pm\frac{k\pi}{2}} \xi_k(v) + e^{\mp\frac{k\pi}{2}} \xi_{-k}(v) \right]
\end{equation}
Define now the lefthand/righthand moving components of $e_k(b)$
\begin{equation}
  e^{\pm}_k (b) := \frac{N_k}{2} e^{\pm x(b)} \ .
\end{equation}
Their transform into $v$ representation is given in terms of $\xi$ functions
\begin{equation}
  [\mathcal{F}^{-1} e^{\pm}_k](v) = \frac{\sqrt{|v|}}{2\pi}N_k \xi_{\pm k}(v) 
\end{equation}
which after plugging back into \eqref{eq:proj_ev}, substituting the value of $N_k$ via \eqref{eq:LQC-normalization} and switching back to $b$-representation gives
\begin{equation}\label{eq:proj-brep}
  [P^{\pm} e_k](b) = \frac{\sinh(k\pi/2)}{\sinh(k\pi)} 
  \left[ e^{\pm\frac{k\pi}{2}} e^{+}_k(b) + e^{\mp\frac{k\pi}{2}} e^{-}_k(b) \right] 
  = e^{\pm}_k(b) + O(\frac{e^{-k\pi}}{\sqrt{k}}) 
  \ .
\end{equation}
In consequence, for large $k$ the projection onto positive/negative $v$ corresponds (to great precision) to the projection onto lefthand/righthand moving plane waves in $x(b)$ variable.

\section{Central moments of semiclassical states}
\label{sec:central}

Consider a 1-dimensional quantum system admitting a pair of fundamental observables being quantum counterparts of classical canonical ones and forming a Heisenberg algebra $[\hat{X},\hat{P}]= i\hbar\mathbb{I}$. Their expectation values form a pair of coordinates\footnote{More precisely, the classes of equivalence of states with respect to relation $\ket{\Psi_1}\sim\ket{\Psi_2} \Leftrightarrow \ket{\Psi_1}=\lambda\ket{\Psi_2}, \lambda\in \mathbb{C}\setminus \{0\}$ -- the so-called \emph{rays} form a manifold \cite{Ashtekar:1997ud} equipped with metric induced by the Hilbert space scalar product. One can then consider coordinates on that manifold.} on the space of quantum states, though this set is obviously incomplete -- one cannot reproduce the state just by knowing their values. One of the possible ways of completing the coordinate system originates from the \emph{Hamburger decomposition} and is based on the construction of the so-called \emph{central moments} \cite{Bojowald:2010qm,Bojowald:2005cw} defined per analogy with such of statistical mechanics $G^{ab}={}"\langle(X-\bar{X})^a(P-\bar{P})^b\rangle{}"$, namely
\begin{equation}\label{gab1}\begin{split}
  G^{a,b}
  &:=\sum_{k=0}^a\sum_{n=0}^b(-1)^{a+b-k-n}\binom{a}{k}\binom{b}{n}\braket{P}^{a-k}\braket{X}^{b-n}\braket{\hat{P}^k\hat{X}^n}_{\rm Weyl} \\
  &= {}" \braket{(\hat{P}-\braket{P})^a(\hat{X}-\braket{X})^b} {}" \ ,
\end{split}\end{equation}
where the last expression is the intuition of the definition, out of which the precise definition (the middle expression) is inferred by a binomial decomposition. All the operators are ordered in the completely symmetric (Weyl) factor ordering. Together with the expectation values $X:=\braket{\hat{X}}, P:=\braket{\hat{P}}$ they form a Poisson algebra, with Poisson brackets uniquely determined by the conditions
\begin{equation}\label{eq:centr-inv}
  \{\braket{\hat{F}^{a,b}},\braket{\hat{F}^{c,d}}\} = -\frac{i}{\hbar} \braket{[\hat{F}^{a,b},\hat{F}^{c,d}]} \ , \qquad
  \hat{F}^{a,b}:=(\hat{X}^a\hat{Y}^b)_{\rm Weyl} \ .
\end{equation}
A huge advantage of such coordinate system is that it admits a hierarchy of quantum corrections with respect to the order $n=a+b$. In particular the 1st order moments always vanish and the 2nd order ones $\{G^{20}, G^{02}, G^{11}\}$ correspond to the state's variance in $X$, $P$ and their correlation respecetively.
In particular, this set allows for a precise definition of semiclassical states as those for which the values of $G^{ab}$ decay sufficiently fast with the order. 

Remarkably, the expectation value of each sufficiently well-behaved composite observable $\hat{O}:=f(\hat{X},\hat{P})$ can be decomposed in terms of the above moments
\begin{equation}\label{eq:O-dec}
  O(X.P,G^{a,b}) := \braket{f(\hat{X},\hat{P})}_{\rm Weyl} 
  = \sum_{a,b=0}^{\infty} \frac{1}{a!b!} \frac{\partial^{a+b} f}{\partial^a X \partial^b P} G^{a,b} \ , 
\end{equation}
where $G^{00}=1$ and $G^{10}=G^{01}=0$. This applies in particular to the Hamiltonian, thus allowing to determine a full set of equations of motion for the variables $\{X,P.G^{ab}\}$ as Hamilton's equations.

In practical applications one often needs to express in terms of central moments an expectation value of some operator, that is not necessarily Weyl-ordered.  In that, it is useful to have some transformation formulas. We recall here a few of such following \cite{article} and applied to operators of the type $P^kX^l$:
\begin{itemize}
  \item a transformation from arbitrary to Weyl factor ordering
    \begin{equation}
      \braket{\hat{P}^k\hat{X}^n}_{x}=\sum_{j=0}^{\min(k,n)}a_j\frac{k!}{(k-j)!}\frac{n!}{(n-j)!}\braket{\hat{P}^{k-j}\hat{X}^{n-j}}_{\rm Weyl} \ ,
    \end{equation}
    where $a_j$ are some coefficients depending on the ordering rule used in $\braket{\cdot|\cdot}_x$, evaluated as coefficients in the expansion of the kernel of the generalized Weyl transform \cite{Cohen-Weyl}.
  \item a transformation from a two-term symmetric factor ordering $\braket{\cdot|\cdot}_s$ (for which the only nonvanishing coefficients equal $a_{2j}= (i/2)^{2j}/(2j)!$, see sec.~IV.A of \cite{article} for the details) 
    \begin{equation}
      \braket{\hat{P}^k\hat{X}^n}_{s} 
      := \frac{1}{2} \braket{P^kX^n+X^nP^k} 
      = \sum_{j=0}^{\min(k,n)}\frac{(i\hbar/2)^{2j}}{(2j)!}\frac{k!}{(k-2j)!}\frac{n!}{(n-2j)!}\braket{\hat{P}^{k-2j}\hat{X}^{n-2j}}_{\rm Weyl} \ .
    \label{simple}
    \end{equation}
\end{itemize}
Finally the Weyl-ordered monomials of $(P,X)$ can be expressed in terms of $(X,P,G^{ab})$ via an inverse inverse of \eqref{gab1}
\begin{equation}
  \braket{\hat{P}^k\hat{X}^n}_{\rm Weyl}=\sum_{i=0}^k\sum_{j=0}^n\binom{n}{j}\binom{k}{i}\braket{P}^{k-i}\braket{X}^{n-j}G^{i,j} \ .
  \label{gab2}
\end{equation}

\section{Trajectory of WDW analog in volume representation}
\label{sec:WDW-traj-v}

A variation of the method introduced in sec.~\ref{sec:trajectories} can be used to evaluate the quantum trajectory of the volume $V(\phi)$ (and its variance) for the Wheeler-DeWitt analog of the studied model in $v$-representation directly. To do so, we note that on the sector of symmetric states the scalar product can be evaluated on the domain $v>0$ only
\begin{equation}
  \braket{\Psi|\Psi'} = 2\int_{\re^+} \rd v \bar{\Psi}(v)\Psi'(v) \ .
\end{equation}
Here, it is again convenient to work with the energy eigenbasis \eqref{eq:WDW-basis}.
A semiclassical state sharply peaked at large (positive or negative) $|k|\gg 1$ can be approximated by  
\begin{equation}
  \Psi^{\sigma}(v,\phi) 
  = \int_{\re}\rd k \Psi(k)\theta(-\sigma k) \frac{1}{\sqrt{4\pi|v|}} e^{ik(x-\sigma\beta\phi)} \ ,
\end{equation}
where $\sigma=1$ for the states peaked about $k_0<0$ and $-1$ for these peaked about $k>0$.

Setting $x:=\ln(v)$ we can write the expectation value of $\hat{V}_{\phi}$ as
\begin{equation}
  \braket{\hat{V}_{\phi}} 
  = \frac{2\alpha}{4\pi} \int_{\re} \rd x \int_{\re^2} \rd k \rd k' \bar{\Psi}^\sigma(k') e^{-ik'(x-\sigma\beta\phi)} \exp(x) \Psi^\sigma(k) e^{ik(x-\sigma\beta\phi)} \ ,
\end{equation}
where $\Psi^\sigma(k) = \Psi(k)\theta(-k\sigma)$. By Taylor expanding $e^x$ in $x=0$, substituting the operators $\hat{x}^n = (i\partial_k+\sigma\beta\phi)$, expanding the powers by binomial theorem we get \begin{equation}
  \braket{\hat{V}_{\phi}} 
  = \alpha \int_{\re} \rd k \bar{\Psi}^\sigma(k) \sum_{n=0}^{\infty} \sum_{j=0}^{n} \frac{1}{n!} \binom{n}{j} (\sigma\beta\phi)^{n-j} (i\partial_k)^j \Psi^\sigma(k) 
  = \alpha \sum_{n=0}^{\infty} \sum_{j=0}^{n} \frac{1}{n!} \binom{n}{j} (\sigma\beta\phi)^{n-j} \braket{(i\partial_k)^j} \ .
\end{equation}
Now, introducing (auxiliary) central moments
\begin{equation}
  \tilde{G}^{0,l} 
  = \sum_{j=0}^l \binom{l}{j} (-\braket{i\partial_k})^{l-j} \braket{(i\partial_k)^j} 
  = "\braket{(i\partial_k-\braket{i\partial_k}\mathbb{I})^l}"  \ ,
\end{equation}
(where again, the last expression represents an intuition), reexpressing the terms $\braket{(i\partial_k)^j}$ by them (using the transformation formula \eqref{eq:centr-inv}) and reordering the sums we arrive to the following form of the expectation value
\begin{equation}\label{eq:WDW-V-v}\begin{split}
  \braket{\hat{V}_{\phi}} 
  &= \alpha\sum_{l=0}^{\infty} \sum_{n=l}^{\infty} \sum_{j=l}^n \frac{1}{n!}\binom{n}{j}\binom{j}{l} (\sigma\beta\phi)^{n-j}\braket{i\partial_k}^{j-l} \tilde{G}^{0,l} 
  = \alpha\sum_{l=0}^{\infty} \sum_{n=0}^{\infty} \sum_{j=0}^n \frac{1}{n!l!}\binom{n}{j} (\sigma\beta\phi)^{n-j} \braket{i\partial_k}^j \tilde{G}^{0,l}  \\
  &= \alpha\exp\left(\sigma\beta\phi + \braket{i\partial_k}\right)\sum_{l=0}^\infty \frac{1}{l!} \tilde{G}^{0,l} \ .
\end{split}\end{equation}
Similarly, one can evaluate the expectation value of squared volume at a given $\phi$
\begin{equation}\begin{split}
  \braket{\hat{V}^2_{\phi}} 
  &= \frac{2\alpha^2}{4\pi} \int_{\re} \rd x \int_{\re^2} \rd k \rd k' \bar{\Psi}^\sigma(k') e^{-ik'(x-\sigma\beta\phi)} \exp(2x) \Psi^\sigma(k) e^{ik(x-\sigma\beta\phi)} \\
  &= \alpha^2\exp\big(2(\sigma\beta\phi + \braket{i\partial_k})\big)\sum_{l=0}^\infty \frac{2^l}{l!} \tilde{G}^{0,l} \ ,
\end{split}\end{equation}
(which is performed exactly as the derivation of \eqref{eq:WDW-V-v} with the only changes $\alpha\mapsto\alpha^2$ and $\exp(x)\mapsto\exp(2x)$) which in turn allows to evaluate the square of variance
\begin{equation}
  \sigma^2(\hat{V}_{\phi})
  = \braket{\hat{V}^2_{\phi}} - \braket{\hat{V}_{\phi}}^2 
  = \alpha^2 \exp\big(2(\sigma\beta\phi + \braket{i\partial_k})\big) \sum_{l=0}^\infty \left(
    \frac{2^l}{l!} \tilde{G}^{0,l} - \frac{1}{l!}\sum_{n=0}^l \binom{l}{n} \tilde{G}^{0,n}\tilde{G}^{0,l-n}
    \right) \ ,
\end{equation}
where in the 2nd term we reordered the summation to group all the components by their total order. Note that the terms for $l=0,1$ vanish, thus the 2nd order term is (as expected) the first nontrivial one. 

In order to bring these results to a more physically meaningful form, in the last step we perform the change of variables the same way, as done in \eqref{eq:var-WDW}. Thus 
\begin{subequations}\label{eq:WDW-VdV-v}\begin{align}
  \braket{\hat{V}_{\phi}} 
  &= \alpha\exp\left(\sigma\beta(\phi -\phi_0))\right)\sum_{l=0}^\infty \frac{(-\sigma\beta)^l}{l!} G^{0,l} \\
  \sigma^2(\hat{V}_{\phi})
  &= \alpha^2 \exp\big(2\sigma\beta\phi -\phi_0) \big) \sum_{l=0}^\infty \frac{(-\sigma\beta)^l}{l!} \left(
    2^l G^{0,l} - \sum_{n=0}^l \binom{l}{n} G^{0,n} G^{0,l-n}
    \right) \ .
\end{align}\end{subequations}

\bibliography{kp-prescriptions}

\end{document}